\documentclass[journal,a4paper,twocolumn]{IEEEtran}
\usepackage{cite}
\usepackage{stackrel}
\usepackage{graphicx}
\usepackage{epstopdf}
\usepackage{subfigure}
\usepackage{amsmath}
\usepackage{amsfonts}
\usepackage{xfrac}
\usepackage{algorithmic}
\usepackage{array}
\usepackage{stfloats}
\usepackage{multirow}
\usepackage{color}
\usepackage{tabulary}
\usepackage[usenames,dvipsnames]{xcolor}
\usepackage{colortbl}
\usepackage{pbox}
\usepackage{booktabs}
\usepackage{amssymb}
\usepackage{setspace}
\usepackage{bigints}
\usepackage{float}
\usepackage[T1]{fontenc}

\providecommand{\norm}[1]{\lVert#1\rVert}
\begin{document}
\newcounter{MYtempeqncnt}
\title{Flexible Dual-Connectivity Spectrum Aggregation for Decoupled Uplink and Downlink Access\\in 5G Heterogeneous Systems}
	
\author{\IEEEauthorblockN{Maria A. Lema\IEEEauthorrefmark{1}, Enric Pardo\IEEEauthorrefmark{1}, Olga Galinina\IEEEauthorrefmark{4}, Sergey Andreev\IEEEauthorrefmark{4}, and Mischa Dohler\IEEEauthorrefmark{1},~\IEEEmembership{Fellow,~IEEE}}\\
\IEEEauthorblockA{\IEEEauthorrefmark{1}Centre for Telecommunications Research, Department of Informatics\\
King's College London,\\}
\IEEEauthorblockA{\IEEEauthorrefmark{4}Department of Electronics and Communications Engineering\\
	Tampere University of Technology\\\{maria.lema\_rosas; enric.pardo; mischa.dohler@kcl.ac.uk; olga.galinina; sergey.andreev@tut.fi\}}}

\maketitle
		
\begin{abstract}
Maintaining multiple wireless connections is a promising solution to boost capacity in fifth-generation (5G) networks, where user equipment is able to consume radio resources of several serving cells simultaneously and potentially aggregate bandwidth across all of them. The emerging dual connectivity paradigm can be regarded as an attractive access mechanism in dense heterogeneous 5G networks, where bandwidth sharing and cooperative techniques are evolving to meet the increased capacity requirements. Dual connectivity in the uplink remained highly controversial, since the user device has a limited power budget to share between two different access points, especially when located close to the cell edge. On the other hand, in an attempt to enhance the uplink communications performance, the concept of uplink and downlink decoupling has recently been introduced. Leveraging these latest developments, our work significantly advances prior art by proposing and investigating the concept of flexible cell association in dual connectivity scenarios, where users are able to aggregate resources from more than one serving cell. In this setup, the preferred association policies for the uplink may differ from those for the downlink, thereby allowing for a truly decoupled access. With the use of stochastic geometry, the dual connectivity association regions for decoupled access are derived and the resultant performance is evaluated in terms of capacity gains over the conventional downlink received power access policies.
\end{abstract}
	
\IEEEpeerreviewmaketitle
\section{Introduction}
The deployment of a practical high-rate network with adequate spectral efficiency requires a variety of innovative features, since contemporary link-level solutions have evolved to approach the Shannon limit with the use of advanced Modulation and Coding Schemes (MCS) \cite{sesia2009lte}. With the goal of improving the per-user throughput and overall system capacity, the Third Generation Partnership Project (3GPP) has recently introduced the concept of Dual-Connectivity in Heterogeneous Networks (HetNets) in the Release 12 \cite{3gpp.36.842}, which is defined as the simultaneous use of spectrum from macro and small cells (MCell and SCell) connected via non-ideal backhaul links over the X2 interface. In this regard, Dual-Connectivity constitutes a novel feature that contributes to achieving the bandwidth demand and improving the data rates by enabling  the User Equipment (UE) to maintain two simultaneous connections. More broadly, multi-connectivity solutions allow to improve the user session continuity by enhancing user connectivity experience as well as the overall communications reliability.

In general, spectrum aggregation techniques are almost directly applicable in the downlink (DL), where power availability to meet the increased bandwidth allocations is not an issue, given that the evolved Node B (eNB) is in charge of radio transmissions. The more restrictive link is often the uplink (UL), as it relies on the user device for carrying out the transmission procedures. Accordingly, an extension in the allocated bandwidth may not be as beneficial in light of the UE power limitations. In the past, similar reasoning was used in the context of Carrier Aggregation, and multiple research works studied and evaluated the feasibility of spectrum aggregation for the UL transmissions \cite{C1,J1}. In connection to Dual-Connectivity, keeping more than one UL connection can be less power efficient for users that are located near the cell edge \cite{R2-140054, AsDude}, primarily due to the increased path-loss experienced toward the serving cells.

As a separate effort, the 3GPP introduces a notion of the UL and DL split in \cite{3gpp.36.842} to offload the MCell more efficiently as well as improve the UL performance. As a consequence of severe transmit power disparities among macro and small cells, the cell that provides the best received power in the DL may not be the same that receives the highest power in the UL. The conventional cell association schemes are based on the DL received power and may result in highly sub-optimal association performance for the UL. Hence, allowing for novel cell association rules in heterogeneous deployments, where energy savings and improved user satisfaction along the cell radius are pursued, can contribute to achieving the UL rates with a higher degree of fairness.

The increased flexibility made available with decoupled UL and DL associations offers advantages when selecting the UL and DL cooperative transmission or reception with the use of Dual-Connectivity. This flexible association, supported by the interoperability of the Downlink and Uplink Decoupling (DUDe) with Dual-Connectivity, makes a decisive step forward in improved multi-connectivity networking. This is because the UE can choose the number and the locations of its DL and UL serving cells independently, and in agreement with several important considerations, such as backhaul capacity, power limitations, and throughput performance, among others. In this sense, spectrum aggregation with the use of Dual-Connectivity becomes more efficient and flexible, thus allowing to maximize the user spectral efficiency.

This work addresses the challenge of efficient cell association in a HetNet system, where users are allowed to aggregate bandwidth with the use of Dual-Connectivity. With the goal of improving the UL capacity and spectral efficiency of aggregated transmissions, the UE follows a per-link maximum received power association rule, which enables to potentially decouple both links. Adding this extra level of flexibility in a multi-connectivity network, exploits all the available benefits of decoupled associations and enhances both UL and DL communications performance.

The rest of this text is organized as follows. This section continues with a literature overview that covers the prior art in both multi-site spectrum aggregation and decoupled associations, and closes with summarizing the main contributions of this work. Section II describes the system model and its underlying assumptions for the stochastic geometry based analysis. In Section III, the association regions and probabilities are derived, and Section IV develops the decoupled capacity expressions. Performance evaluation is conducted in Section V, followed by the conclusions.

\subsection{State of the Art Review}
Dual connectivity was proposed by the 3GPP as an architectural solution to improve user performance by combining the benefits of the MCell coverage and the SCell capacity, where the Release 10 Carrier Aggregation is applied to aggregate carriers in co-channel HetNets. The potential of this technology has been widely studied by the 3GPP in \cite{3gpp.36.842}, where significant capacity gains were recognized.

Further, spectrum aggregation techniques were well-studied for the DL, ever since the introduction of Carrier Aggregation in 3GPP. Several works demonstrated available improvements in the DL throughput brought by aggregating the transmissions, which covered such topics as: (a) user distribution among different carriers to perform load balancing \cite{11}, (b) carrier selection strategies that consider various frequencies and hence dissimilar coverage footprints \cite{25}, and (c) carrier aggregated scheduling procedures \cite{2}. Moreover, the research community thoroughly investigated the performance improvements made available with the use of inter-site resource aggregation supported by the Carrier Aggregation.

The work in \cite{IS-CA1} studied the inter-site aggregation in a DL scenario, where MCells share resources with other cells. The paper in question proposed a Carrier Aggregation window to determine whether the Carrier Aggregation-compliant UEs should be selected to consume resources from both cells. This work considered a dedicated frequency deployment, where all cells are assumed to be operating on different frequencies. The benefits of aggregating resources across both cells were verified for various traffic patterns and cell load situations. Inter-site Carrier Aggregation has also been applied as a solution to enhance mobility and handover procedures in HetNets: a primary connection is always kept with the MCell and when the UE is inside the coverage of a SCell, it configures a secondary carrier and can thereby enjoy the spectrum aggregation benefits \cite{6515048}.

In the context of Dual-Connectivity, several research papers focused on the many open challenges and analyzed the potential performance improvements. The work in \cite{7393786} tackled the DL scheduling aspects and proposed a downlink traffic scheduling mechanism that aims at maximizing the network throughput when deciding on the traffic split to the SCell. Further, the work in \cite{6825019} studied the Dual-Connectivity with a Control/User Plane split and proposed a flexible network configuration, which employs the channel state information reference signal (CSI-RS) knowledge for SCell association purposes. Similarly, the authors in \cite{7405721} addressed the association as an optimization problem: the optimal combination of macro and small cells and the optimal traffic split between both serving cells were investigated. The improvements in user performance with the utilization of shared resources provided a strong indication that cooperative techniques are becoming essential to maximize both spectrum utilization and efficiency. A comprehensive overview of the architectural enhancements together with the detailed performance results is available in \cite{7498101}.

While various spectrum aggregation techniques in the DL were explored  widely, the UL has remained much more controversial. This is because the gains in the UL are less straightforward, since the physical layer issues such as power de-rating need to be considered when assessing user eligibility for multiple transmissions \cite{C1,J1}. Some works showed clearly the potential of spectrum aggregation in the UL: inter-site Carrier Aggregation was explored in \cite{IS-CA2}, where the results indicated improved UL throughput levels in the low load situations due to a larger bandwidth accessibility. Similarly, the work in \cite{6514952} demonstrated visible benefits for the UL when considering multi-carrier transmissions. Based on these findings, multiple connections may indeed help improve the overall UL performance and boost the cell-edge transmissions. In this context, the expected gains are twofold: the UE can improve its throughput by accessing a larger bandwidth as well as due to a better coverage. For Dual-Connectivity, certain differences emerge when compared to Carrier Aggregation. Since resource allocation is performed independently and without an explicit need for coordination, different power scaling and power splitting techniques were proposed in the literature to avoid exceeding the UE maximum transmit power. For instance, enhanced UL power control schemes for Dual-Connectivity were investigated in \cite{6965872}, \cite{7023776}, and \cite{7504389} by means of system level simulations.

One of the most challenging aspects of Dual-Connectivity is the UE power handling, which was widely studied in the context of HetNet-centric UL and DL association rules. One step ahead in the optimization of HetNets is accounting for the relationship between the UL and the DL as well as understanding how the association policies affect the performance. The UL/DL power together with the MCell/SCell load and power imbalance motivate the decoupling of both links, which is deemed particularly beneficial in co-channel heterogeneous deployments. In the Release~12, the 3GPP provided an initial evaluation of the HetNet performance when including the UL and DL split. These results confirmed improvements especially at the cell edge for both low and medium load scenarios \cite{3gpp.36.842, R2-131678}. The research literature tackled the power and load imbalance challenges recently, and a number of relevant references could be identified as follows.

The authors in \cite{DUDe} developed a path-loss cell association solution to the power imbalance problem. The results in terms of gains that could be achieved in the UL capacity are very promising. A detailed analysis of the decoupled access in terms of the association probability, coverage, and capacity was reported in \cite{AsDude}. There, the prior work was extended by adding an analytical evaluation based on stochastic geometry and architectural considerations. The findings revealed similar trends between the stochastic geometry analysis and the real-world experimental data. The work in \cite{LoadDUDe} introduced the cell load and the backhaul limitations into the cell association process. The signal-to-interference-plus-noise ratio (SINR) variance was reduced with the proposed DUDe solution. In addition, the applied interference-aware UL power control enabled further improvements in the UL throughput. Finally, \cite{PLDUDe} analyzed the UL SINR and the rate distributions as functions of the association rules accounting for the UL power control design parameters. The results demonstrated that the minimum path-loss association leads to the identical load distribution across all cells that also remains optimal in terms of rate, irrespective of the power control parameters. As an outcome, when both the UL and DL joint coverage has to be maximized, the decoupled association is the optimal choice. This is because it reduces the QoS imbalance between both links.

In summary, considering a Dual-Connectivity scenario jointly with a decoupled association policy appears to be an attractive solution to address many of the UL spectrum aggregation challenges at the cell edge, as well as unlock the available benefits of extending the bandwidth accessibility.
\subsection{Main Contributions}
This work significantly extends the prior art regarding the decoupled association in \cite{DUDe,AsDude,PLDUDe} by proposing and carefully investigating the concept of flexible cell association in the Dual-Connectivity scenarios, where users are allowed to aggregate spectrum thus boosting their capacity. Given that the decoupled associations have been proposed to improve the UL communications performance, our study focuses on analyzing the respective benefits for this link specifically. The Dual-Connectivity investigation for the DL is therefore left out of scope of this work.

In a Dual-Connectivity scenario, we consider two uplink associations. The user is allowed to attach to the first and the second best serving stations by following the UL received power policies. In such conditions, we evaluate whether decoupled associations offer improvements with respect to the conventional downlink received power association rules. In addition, the feasibility of maintaining more than one connection in the UL is discussed, since UE power limitations may impair the throughput performance. The main contributions of this work can be summarized as follows:
\begin{itemize}
	\item systematic recognition of different user association cases that reflect the Dual-Connectivity aggregated transmissions;
	\item rigorous stochastic geometry based modeling of a two tier co-channel HetNet with flexible associations in the Dual-Connectivity context; and
	\item comprehensive mathematical analysis and derivation of the association probabilities as well as the capacity performance metrics for the Dual-Connectivity aggregated transmissions.
\end{itemize}

\section{Modeling the Dual Connectivity Association} \label{system}
In this section, we introduce the main system modeling assumptions as well as discuss their suitability and relevance.	%

\subsubsection{Infrastructure deployment}
We assume for simplicity that the MCell and SCell locations on a plane are independent and follow a homogeneous Poisson Point Process (PPP) with intensities $\lambda_m$ and $\lambda_s$, respectively ($\lambda_s\!>\!\lambda_m$). The area of a MCell is defined as the locus of points, which are geometrically closer to the selected MCell than to any other MCell on the plane, and represented by a Voronoi cell.

We note that the PPP is a commonly used and mathematically tractable formulation that captures randomness of the SCells locations. In reality, the MCells may be distributed in a more deterministic manner and could also be characterized by e.g., repulsion point processes. However, in our modeling  such determinism would only alter the probability distribution of distances between the user and the MCells, without any impact on the generality of our proposed methodology.

Importantly, if the intensity of the SCells $ \lambda_s$ is considerably higher than that of the MCells $ \lambda_m$, then several SCells may be located within the area of one MCell.

\subsubsection{User locations}
The UE locations are assumed to be static and distributed according to the homogeneous PPP with the intensity $\lambda_d$. For the purposes of its spatial statistics characterization, the HetNet in question is regarded as a single snapshot in time.

\subsubsection{User associations}
Cell association policies defined by the 3GPP are based on the DL reference signal received power (RSRP), which provides information on the signal strength and offers no indication of its quality. When allowing for decoupled associations, a per-link association policy is applied. Hence, similar to the DL, the UL received power is considered when selecting the serving base station. The main rationale behind concentrating on the UE signal strength that includes no information on the experienced interference is in the fact that the interference may change rapidly and abruptly, as well as can suddenly rule out the prior selection based on the instantaneous measurements.

\subsubsection{Signal propagation}
The received UL/DL signal, when averaged over time, is assumed to follow the dependence on the distance between the transmitter and the receiver in the form of:

\begin{equation}
E_h[{S}] = P_{tx} \ \norm{X}^{-\alpha},  \label{Sv DL}
\end{equation}
where $E_h[{S}]$ is the average UL/DL signal received power (index $h$ stands for averaging over the Rayleigh fading variations based on the exponentially distributed random variable with a unit mean), $P_{tx}$ is the transmit power including the antenna gain, $\norm{X}$ is the distance from the UE to the serving station, and ${\alpha}$ is the path-loss exponent.

This elegant approximation of the signal propagation corresponds to the contemporary 3GPP considerations, which provide the coefficient $\alpha$ as well as an additional linear multiplier that is included here for clarity into the value of $P_{tx}$.

\subsubsection{Power control}
The transmit power of the MCell and SCell transmitters is constant and equals to $P_m$ and $P_s$, respectively. 
	%
	%

The transmit power of the UE $P_d$ ($P_d<P_s<P_m$) follows the rules of the open-loop power control (OLPC) mechanism defined in \cite{3gpp.36.213}. Accordingly, the user establishes an operating point using the open-loop procedure, where it compensates for the mean path-loss and its slow variations. Hence, the user transmit power may be expressed as:
\begin{equation}
P_d=P_{\text{0}} L_{x}^\gamma, \label{OLPC_Stoch}
\end{equation}
where $L_{x}$ corresponds to the distance-dependent user path-loss; $P_{\text{0}}$ and $\gamma$ are the OLPC parameters.

For simplicity of our further analysis, we consider the constant UE transmit power $P_d$ as an example, so that there is no path-loss compensation. However, our subsequent derivations could easily be extended for the case of variable $L_{x}^\gamma$. Further, the user device is allowed to only make an equal power splitting between the two serving cells; hence, the total transmit power budget is divided into two components. As a result, every user will transmit to each of the serving stations with a half of its maximum power $\frac{P_d}{2}$. Following the work in \cite{6965872}, we note that a path-loss based power splitting model may impair the performance of the small cell, whereas the considered equal power splitting policy performs near similar to enhanced UL power control for dual-connectivity.

\subsubsection{UL capacity}
We assume that the UL capacity can be produced based on the Shannon formula:
\begin{align}
&C ={B} \log_2 \left(1 + \frac{P_{rx}}{I_x } \right),
\end{align}
where $B$ is the available bandwidth, 
$P_{rx}$ is the received signal power, and $I_x$ is the interference term (not including the noise component). This formula, although widely-adopted in its current form, may be adjusted further with the linear multipliers and thus perfectly match the 3GPP calibration data (see e.g., \cite{galinina2013energy} for details). 

\subsubsection{Interference model}
The HetNet in question is assumed to be interference limited, that is, the interference from other users dominates the noise power. As it is typical for the UL in 3GPP systems, we do not consider intra-cell interference and focus exclusively on the inter-cell interference. The corresponding interference model largely follows the approach of \cite{UL}: we assume that there is a single dominant interference source per cell. As a result, the number of interfering users equals the number of cells and all of the interferes are located outside of the cell (i.e., farther away than the tagged UE).
	
\subsubsection{Target scenario}
In what follows, we consider one \textit{cluster} formed by a single MCell and two SCells as a representative construction block that composes practical multi-cell scenarios. In particular, we focus on a tagged UE, as shown in Fig.~\ref{fig:all_cases}.
The UE under consideration is connected simultaneously to the two serving cells, and the set of possible associations includes three options: the corresponding MCell, the closest SCell, and the second closest SCell. As it is widely known, the distance to the closest point of a PPP (in our example, the distance to the associated MCell) follows a Rayleigh distribution with the probability density function (pdf) of $f_x(x) = 2\lambda \pi x e^{-\lambda \pi x^2}$. However, the distribution of distances to the $n$th-closest point drawn from a PPP realization might lead to cumbersome derivations, and thus for the sake of exposition we approximate the distributions of distances to the closest SCells by two independent Rayleigh distributions.


\section{Association Probability for Decoupled Access}
The primary goal of this study is to establish the probability of the decoupled events while having simultaneous UL and DL connections to \textit{two} serving stations. As explained in \cite{AsDude}, there are some decoupled association combinations that are not possible under the present assumptions: in particular, the first DL connection to a SCell and the first UL connection to a MCell due to the fact that $P_m>P_s$. In the following subsections, we demonstrate that this association case is only possible for the second connection.
	
Based on the above assumptions, there are in total \textit{six} association possibilities for the model under consideration. These six association cases are further split into \textit{twelve} subcases, a summary of which is provided in Table~\ref{table1}. Our mathematical derivation of the association probability is delivered in what follows.

\begin{table*}[tbp]
\caption{Probability regions in the best association Dual-Connectivity scenario}\label{table1}
		\centering
		\begin{tabular}{c c c c c c c}
			\toprule
			 \textbf{\footnotesize Case}       & \textbf{\footnotesize Subcase} & \textbf{\footnotesize UL 1st } & \textbf{\footnotesize DL 1st} & \textbf{\footnotesize UL 2nd } & \textbf{\footnotesize DL 2nd }    &     \textbf{\footnotesize Restricting inequalities }\\ \midrule
			
			   & \footnotesize 1.1	& \footnotesize MCell        & \footnotesize MCell        & \footnotesize SCell 1	& \footnotesize SCell 1	&${x_m}< {x_1}< {x_2}$
\\ \cmidrule{2-6}
			
\multirow{-2}{*}{1} & \footnotesize  1.2	& \footnotesize MCell        & \footnotesize MCell        & \footnotesize SCell 2	& \footnotesize SCell 2	&${x_m}< {x_2}< {x_1}$   
\\ \midrule

& \footnotesize 2.1	& \footnotesize SCell 1      & \footnotesize SCell 1      & \footnotesize MCell	&  \footnotesize MCell		& $\frac{P_m}{P_s}{x_1} <{x_m}<{x_2} $   
			  \\ \cmidrule{2-6}
			
\multirow{-2}{*}{2} & \footnotesize 2.2	& \footnotesize SCell 2      & \footnotesize SCell 2      & \footnotesize MCell	& \footnotesize MCell		& $\frac{P_m}{P_s}{x_2}< {x_m}<{x_1}$      
\\ \midrule
					
			   &\footnotesize 3.1	&\footnotesize SCell 1      &\footnotesize SCell 1      &\footnotesize SCell 2	&\footnotesize MCell		&${x_1}< \left(\frac{P_s}{P_m}\right)^{1/\alpha}  {x_m}<  {x_2}<  {x_m}$
			   \\ \cmidrule{2-6}
			
\multirow{-2}{*}{3} & \footnotesize 3.2	& \footnotesize SCell 2      & \footnotesize SCell 2      &  \footnotesize SCell 1	& \footnotesize MCell		& ${x_2}< \left(\frac{P_s}{P_m}\right)^{1/\alpha}  {x_m}<  {x_1}<  {x_m}$
\\ \midrule

 &\footnotesize 4.1	& \footnotesize SCell 1      &\footnotesize MCell      &\footnotesize SCell 2	&\footnotesize SCell 1		&${x_1}<{x_2}<  {x_m}<\left(\frac{P_m}{P_s}\right)^{1/\alpha}  {x_1}$
 \\ \cmidrule{2-6}

 \multirow{-2}{*}{4}  &\footnotesize 4.2	&\footnotesize SCell 2      &\footnotesize MCell      &\footnotesize SCell 1	&\footnotesize SCell 2		&  ${x_2}<{x_1}<  {x_m}<\left(\frac{P_m}{P_s}\right)^{1/\alpha}  {x_2}$
 \\ \midrule

&\footnotesize 5.1	& \footnotesize SCell 1      &\footnotesize MCell      &\footnotesize MCell	&\footnotesize SCell 1		&${x_m}< {x_2}, {x_1}<x_m< \left(\frac{P_m}{P_s}\right)^{1/\alpha}{x_1}$
\\ \cmidrule{2-6}

\multirow{-2}{*}{5}  &\footnotesize 5.2	&\footnotesize SCell 2      &\footnotesize MCell      &\footnotesize MCell	&\footnotesize SCell 2		& ${x_m}< {x_1}, {x_2}<x_m< \left(\frac{P_m}{P_s}\right)^{1/\alpha}{x_2}$
\\ \midrule	

&\footnotesize 6.1	& \footnotesize SCell 1      &\footnotesize SCell 1      &\footnotesize SCell 2	&\footnotesize SCell 2		&${x_1}<{x_2}< \left(\frac{P_s}{P_m}\right)^{1/\alpha}{x_m}$ \\ \cmidrule{2-6}
			
\multirow{-2}{*}{6}  &\footnotesize 6.2	&\footnotesize SCell 2      &\footnotesize SCell 2      &\footnotesize SCell 1	&\footnotesize SCell 1		&${x_2}<{x_1}<\left( \frac{P_s}{P_m}\right)^{1/\alpha}{x_m}$
\\
\bottomrule
		\end{tabular}
	\end{table*}


\vspace{-5pt}
\subsection{Case 1: Connection with MCell and SCell in UL and DL}
This case considers the probability region, where the first connection is to the MCell and the second one is to either of the SCells in both the UL and the DL. To this end, Fig.~\ref{fig:all_cases} illustrates a graphical representation of this and other association cases.

The consideration at hand leads to the following conditions for subcases 1.1 and 1.2, respectively:
\begin{equation}
	\begin{array}{l r}
	\text{UL}: {P_d}\:\norm{X_m}^{-\alpha} > {P_d}\:\norm{X_1}^{-\alpha} > {P_d}\:\norm{X_2}^{-\alpha} \\  \text{DL}: {P_m}\:\norm{X_m}^{-\alpha} > \  {P_s}\:\norm{X_1}^{-\alpha}  > \  {P_s}\:\norm{X_2}^{-\alpha}
	\end{array}, \label{eqn:ULDL_condition11}
\end{equation}	
\begin{equation}
	\begin{array}{l r}
	\text{UL}: {P_d}\:\norm{X_m}^{-\alpha} > {P_d}\:\norm{X_2}^{-\alpha} > {P_d}\:\norm{X_1}^{-\alpha} \\   \text{DL}: {P_m}\:\norm{X_m}^{-\alpha} > \  {P_s}\:\norm{X_2}^{-\alpha}  > \  {P_s}\:\norm{X_1}^{-\alpha}
	\end{array},\label{eqn:ULDL_condition12} 
\end{equation}
where $\norm {X_1},\norm{X_2}$, and $\norm{X_m}$ are the distances to the SCells and the associated MCell, which we further denote for brevity as $x_1$, $x_2$, and $x_m$, respectively. 	%

In both situations (\ref{eqn:ULDL_condition11}) and (\ref{eqn:ULDL_condition12}), the UL condition is more restrictive. Therefore, the events of this association case can be reduced to:
\begin{equation}
	\begin{array}{c}
 {x_m}^{-\alpha} >  {x_1}^{-\alpha} >  {x_2}^{-\alpha}\Rightarrow {x_m} < \min ({x_1} , {x_2}) . \nonumber
	\end{array}
\end{equation}

Hence, the probability for Case 1 can be calculated as:
\begin{equation}
\begin{array}{c}
	P_{\text{Case 1}}\ = \Pr(x_m<{\min}(x_1,x_2)) = \\
	\int \limits_{0}^{+\infty} (1-F_{\min (x_1,x_2)}(x))\cdot f_{x_m}(x)\ \text{d} x. \label{case 1.1 dc}
\end{array}
\end{equation}

The Cumulative Distribution Function (CDF) of $y=\text{min}(x_1,x_2)$ can be produced by using order statistics, and is given by:
\begin{equation}
	F_{y}(x) = 1-\Pr \left(y>x \right)=1 - (1 - F_{x_1}(x)) (1 - F_{x_2}(x)).   \label{min(x1,x2)}
\end{equation}

Since $F_{x_1} (x)$ and $F_{x_2}(x)$ are assumed to be identical, the expression \eqref{min(x1,x2)} can rearranged as:
\begin{equation}
\begin{array}{c}
	F_{\text{min}(x_1,x_2)}(x) = 1 - (1 - F_{x_1} (x))^2 = 1 - e ^{-2\pi\lambda_sx^2}  \label{min(x1,x2)(xm)}.
\end{array}
\end{equation}

Finally, after substituting \eqref{min(x1,x2)(xm)} into \eqref{case 1.1 dc}, we may simplify the probability for Case 1 down to the following expression:
\begin{equation}
\begin{array}{c}
	P_{\text{Case 1}}\! =\! \int \limits_{0}^{+\infty} e ^{-2\pi\lambda_sx^2} 2\pi\lambda_mx e^{-\pi\lambda_m x^2}\ \!\!\! \text{d} x\!=\! \frac{\lambda_m}{2\lambda_s+\lambda_m}.
\end{array}
\end{equation}

\vspace{-10pt}
\subsection{Case 2: Connection with SCell and MCell in UL and DL}
\begin{figure*}
\centering
\includegraphics[width=1\textwidth]{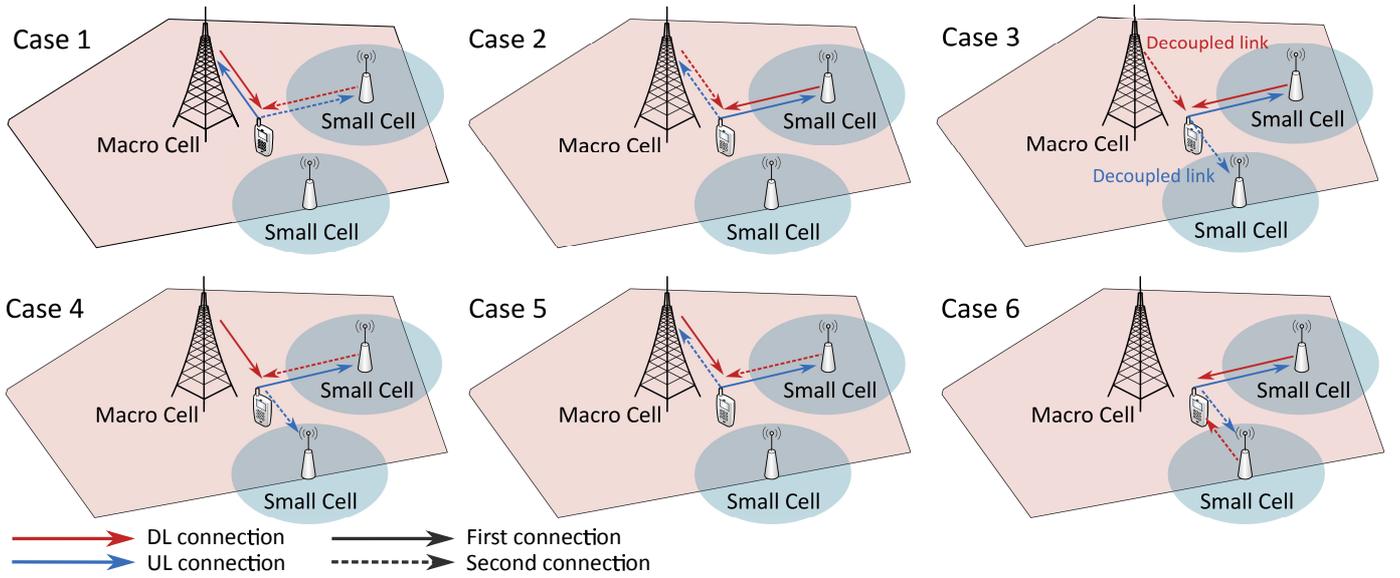}
\caption{The considered association cases.}
\label{fig:all_cases}
\vspace{0.1cm}
\end{figure*}

In Case 2, the connections kept by the UE are similar to those for Case 1, but have another order. The user associates with the SCell (first connection) and then connects to the MCell (second connection), for both the UL and the DL (see Fig.~\ref{fig:all_cases}).

The events related to this case can be expressed as:
\begin{equation}
	\begin{array}{l r}
	\text{UL}: {P_d}\:\norm{X_1}^{-\alpha} > {P_d}\:\norm{X_m}^{-\alpha} > {P_d}\:\norm{X_2}^{-\alpha} \\
	 \text{DL}: {P_s}\:\norm{X_1}^{-\alpha} > \  {P_m}\:\norm{X_m}^{-\alpha}  > \  {P_s}\:\norm{X_2}^{-\alpha}
	\end{array},
\end{equation}
\begin{equation}
	\begin{array}{l r}
	\text{UL}: {P_d}\:\norm{X_2}^{-\alpha} > {P_d}\:\norm{X_m}^{-\alpha} > {P_d}\:\norm{X_1}^{-\alpha} \\
	\text{DL}: {P_s}\:\norm{X_2}^{-\alpha} > \  {P_m}\:\norm{X_m}^{-\alpha}  > \  {P_s}\:\norm{X_1}^{-\alpha}
	\end{array}.
\end{equation}\

Rewriting the above, we arrive at the following conditions for subcases 2.1 and 2.2, respectively:
\begin{equation}
\begin{array}{l r}
	\text{UL}: \: { x_1} < { x_m}<  { x_2} \\
	\text{DL}: \: \left(\frac{P_m}{P_s}\right)^{\frac{1}{\alpha}}{ x_1} <  { x_m}  < \left(\frac{P_m}{P_s}\right)^{\frac{1}{\alpha}}{ x_2}
	\end{array},
\end{equation}
\begin{equation}
	\begin{array}{l r}
	\text{UL}:\: { x_2}< \: { x_m}<\: { x_1} \\
	\text{DL}: \:\left(\frac{P_m}{P_s}\right)^{\frac{1}{\alpha}} { x_2} <{ x_m}  <\left( \frac{P_m}{P_s}\right)^{\frac{1}{\alpha}}{ x_1}
	\end{array}.
\end{equation}

Combining the UL and DL conditions and exploiting the fact that $\frac{P_m}{P_s}>1$, we derive the restricting inequalities that are sufficient for the above to hold:
\begin{equation}
\begin{array}{l}
	\text{Subcase 2.1}:\: \left(\frac{P_m}{P_s}\right)^{\frac{1}{\alpha}}{ x_1} < { x_m}<  { x_2}
	\end{array},
	\label{2.1}
\end{equation}
\begin{equation}
	\begin{array}{l}
	\text{Subcase 2.2}:\: \left(\frac{P_m}{P_s}\right)^{\frac{1}{\alpha}} { x_2}<  { x_m}< { x_1}
	\end{array}. \label{2.2}
\end{equation}

Let us now establish the probability corresponding to \eqref{2.1}: 
\begin{equation}
\begin{array}{l}
	\Pr\left(\left(\frac{P_m}{P_s}\right)^{\frac{1}{\alpha}}{ x_1} < { x_m}<  { x_2} \right) \!\!=\!\!  \int \limits_{0}^{+\infty} f_{x_1}(x_1) \!\! \int \limits_{\left(\frac{P_m}{P_s}\right)^{\frac{1}{\alpha}}x_1}^{+\infty} \!\! f_{x_m}(x_m)
	\\
	 \int \limits_{x_m}^{+\infty}  f_{x_2}(x_2) \text{d}x_2  \text{d}x_m  \text{d}x_1=
	 \frac{ \lambda_s\lambda_m} {(\lambda_s+\lambda_m)(\lambda_s+\eta \lambda_s+\eta \lambda_m)},
\end{array}
\end{equation}
where $\eta = \left(\frac{P_m}{P_s}\right)^{2/\alpha}$.

The latter expression holds for subcase 2.2 with its corresponding restriction \eqref{2.2} (due to the assumption that the distances to the SCells are identically distributed) and thus the
final association probability for Case 2 is:
	\begin{equation}
	P_{\text{Case 2}} = \frac{2 \lambda_s\lambda_m} {(\lambda_s+\lambda_m)(\lambda_s+\eta \lambda_s+\eta \lambda_m)}.
	\end{equation}	
	\vspace{-10pt}

\vspace{-10pt}
\subsection{Case 3: UL and DL connection with SCell, DL to MCell and UL to another SCell}\label{sec:Case3}

This association case is based on a decoupling event. The UE is closer to one SCell, which defines the first association for both the UL and the DL. However, when choosing the second serving cell, the user receives higher power from the MCell, while the UL received power is higher for the second SCell, and hence the user decouples its second connection.

For this scenario, the UL and DL conditions may be derived as follows:
\begin{equation}
\begin{array}{c}
	\!\!\!\!\text{UL}: {P_d}\:\norm{X_1}^{-\alpha} > {P_d}\:\norm{X_2}^{-\alpha} > {P_d}\:\norm{X_m}^{-\alpha} \\
	\text{DL}: {P_s}\:\norm{X_1}^{-\alpha} > \  {P_m}\:\norm{X_m}^{-\alpha}  > \  {P_s}\:\norm{X_2}^{-\alpha}
	\end{array},
\end{equation}
\begin{equation}
\begin{array}{c}
\!\!\!\! \text{UL}: {P_d}\:\norm{X_2}^{-\alpha} > {P_d}\:\norm{X_1}^{-\alpha} > {P_d}\:\norm{X_m}^{-\alpha} \\
 \text{DL}: {P_s}\:\norm{X_2}^{-\alpha} > \  {P_m}\:\norm{X_m}^{-\alpha}  > \  {P_s}\:\norm{X_1}^{-\alpha}
	\end{array},
\end{equation}
with the corresponding restricting conditions:
\begin{equation}
	\text{Subcase 3.1:}\  {x_1}<\frac{x_m}{ \sqrt{\eta}}< {x_2}<  {x_m} \label{3.1},
\end{equation}
\begin{equation}
	\text{Subcase 3.2:}\  {x_2} <\frac{x_m}{ \sqrt{\eta}}<{x_1}<  {x_m} \label{3.2},
\end{equation}
where $\eta = \left(\frac{P_m}{P_s}\right)^{2/\alpha}$.

Let us first consider the situation when $x_1<x_2$. Then, the corresponding probability is:
\begin{equation}
\begin{array}{l}
\Pr({x_1} < \frac{x_m}{ \sqrt{\eta}} <{x_2}  <  {x_m}) = \\
\int \limits_{0}^{+\infty}  f_{x_1}(x_1)  \int \limits_{x_1\sqrt{\eta}}^{+\infty}  f_{x_m}(x_m)
\int \limits_{\frac{x_m}{ \sqrt{\eta}}}^{x_m}  f_{x_2}(x_2) \text{d}x_2  \text{d}x_m  \text{d}x_1=
\\
 \frac{1}{\lambda_m + \lambda_s/\eta}  \frac{\lambda_m\lambda_s}{\lambda_m\eta + 2\lambda_s} - \frac{1}{\lambda_m + \lambda_s}\frac{\lambda_m\lambda_s}{\eta \lambda_m + \lambda_s \eta+\lambda_s}.
\end{array} \nonumber
\end{equation}

We note that for subcase 3.2 the derivations would be similar due to the inherent symmetry. Therefore, we may readily obtain the sought probability for Case 3 as:
	\begin{equation}
	P_{\text{Case 3}} =  \frac{2}{\lambda_m + \frac{\lambda_s}{\eta}}  \frac{\lambda_m\lambda_s}{\lambda_m\eta + 2\lambda_s} - \frac{2}{\lambda_m + \lambda_s}\frac{\lambda_m\lambda_s}{\eta \lambda_m + \lambda_s \eta+\lambda_s},
	\end{equation}
where $\eta = \left(\frac{P_m}{P_s}\right)^{2/\alpha}$.
	

\subsection{Case 4: First UL with SCell and DL with MCell, UL with second SCell and DL with first SCell}	
This case considers two decoupling events: the first connection is decoupled having the first UL associated to the SCell and the DL to the MCell. When choosing the second connection, the best UL is associated with the second closest SCell, since the MCell is farther away. In the DL case, the second connection is best to the closest SCell. This case is particularly interesting since both ULs are connected to different SCells, which is due to the fact that we do not consider the in-cell spectrum aggregation with the use of Carrier Aggregation.

The corresponding UL and DL equations are produced as follows:
\begin{equation}
\begin{array}{c}
\text{UL}: {P_d}\:\norm{X_1}^{-\alpha} > {P_d}\:\norm{X_2}^{-\alpha} > {P_d}\:\norm{X_m}^{-\alpha} \\  \text{DL}: {P_m}\:\norm{X_m}^{-\alpha} > \  {P_s}\:\norm{X_2}^{-\alpha}  > \  {P_s}\:\norm{X_1}^{-\alpha}
\end{array}, \label{4.1}
\end{equation}
\begin{equation}
\begin{array}{c}
\text{UL}: {P_d}\:\norm{X_2}^{-\alpha} > {P_d}\:\norm{X_1}^{-\alpha} > {P_d}\:\norm{X_m}^{-\alpha} \\  \text{DL}: {P_m}\:\norm{X_m}^{-\alpha} > \  {P_s}\:\norm{X_1}^{-\alpha}  > \  {P_s}\:\norm{X_2}^{-\alpha}
\end{array}. \label{4.2}
\end{equation}

From the above, we derive the restrictive conditions for both subcases as:
\begin{equation}
\text{Subcase 4.1:}\ {x_1}<{x_2}< {x_m}< \left(\frac{P_m}{P_s}\right)^{1/\alpha} {x_1},
\end{equation}
\begin{equation}
\text{Subcase 4.2:}\ {x_2}< {x_1}< {x_m}< \left(\frac{P_m}{P_s}\right)^{1/\alpha} {x_2}.
\end{equation}

Let us first consider the situation when $x_1<x_2$. Then, the corresponding probability is:
\begin{equation}
\begin{array}{l}
\Pr( {x_1}<{x_2}< {x_m}< \sqrt{\eta} {x_1}) = \\
\int \limits_{0}^{+\infty}  f_{x_1}(x_1)  \int \limits _{x_1}^{x_1\sqrt{\eta}}  f_{x_2}(x_2)
\int \limits_{x_2}^{x_1\sqrt{\eta}}  f_{x_m}(x_m)   \text{d}x_m  \text{d}x_2 \text{d}x_1=
\\
\frac{\lambda_s^2}{\lambda_m+\lambda_s}\!\!\left(\frac{1}{\lambda_m+2\lambda_s}\!-\!
\frac{1}{\lambda_m \eta+\lambda_s+ \eta \lambda_s} \! \right)\!- \!
\frac{\lambda_s}{\lambda_m \eta+2\lambda_s} \!+\!\frac{\lambda_s}{\lambda_m \eta+\lambda_s \eta+\lambda_s}.
\end{array} \label{eqn:subcase41}
\end{equation}

The result for subcase 4.2 may be obtained by the analogy with the above calculations, and hence the probability for Case 4 can be established as:
\begin{equation}
\begin{array}{c}
P_{\text{Case 4}} =
2\left(\frac{\lambda_s^2}{\lambda_m+\lambda_s}\frac{1}{\lambda_m+2\lambda_s}-
\frac{\lambda_s^2}{\lambda_m +\lambda_s}\frac{1}{\lambda_m \eta+\lambda_s+ \eta \lambda_s}- \right.
\\
\left. \frac{\lambda_s}{\lambda_m \eta+2\lambda_s} +\frac{\lambda_s}{\lambda_m \eta+\lambda_s \eta+\lambda_s} \right).
\end{array} \nonumber
\end{equation}
\vspace{-20pt}
\subsection{Case 5: First UL with SCell and DL with MCell, second UL with MCell and DL with SCell}	
%
In this case, there are also two decoupling events. The UE is closer to one of the SCells as well as to the MCell, while the second SCell is farther away. Based on this, the UE selects its first UL connection to the SCell and the DL is associated with the MCell having as first best connection the decoupled link. When choosing the second connection, the MCell is closer than the SCell. Therefore, the UE observes an inverted decoupled event: the UL to the MCell and the DL to the SCell. This is possible because we do not consider Carrier Aggregation for the association, so that the UE needs to seek for the second best serving cell by excluding the one already selected.

The expressions defining the UL and the DL conditions are:
\begin{equation}
\begin{array}{c}
\text{UL}: {P_d}\:\norm{X_1}^{-\alpha} > {P_d}\:\norm{X_m}^{-\alpha} > {P_d}\:\norm{X_2}^{-\alpha} \\  \text{DL}: {P_m}\:\norm{X_m}^{-\alpha} > \  {P_s}\:\norm{X_1}^{-\alpha}  > \  {P_s}\:\norm{X_2}^{-\alpha}
\end{array}, \label{5.1}
\end{equation}
\begin{equation}
\begin{array}{c}
\text{UL}: {P_d}\:\norm{X_2}^{-\alpha} > {P_d}\:\norm{X_m}^{-\alpha} > {P_d}\:\norm{X_1}^{-\alpha} \\  \text{DL}: {P_m}\:\norm{X_m}^{-\alpha} > \  {P_s}\:\norm{X_2}^{-\alpha}  > \  {P_s}\:\norm{X_1}^{-\alpha}
\end{array}. \label{5.2}
\end{equation}

Simplifying the above, we may write down the following restrictive conditions (based on the DL) for both subcases:
\begin{equation}
\text{Subcase 5.1:}\ {x_m}< {x_2}, {x_1}<x_m< \frac{P_m}{P_s}{x_1},
\end{equation}
\begin{equation}
\text{Subcase 5.2:}\ {x_m}< {x_1}, {x_2}<x_m< \frac{P_m}{P_s}{x_2}.
\end{equation}

Let us consider subcase 5.1 and derive the probability of the event, when ${x_1}<x_m<\sqrt{\eta}{x_1}$:
\begin{equation}
\begin{array}{c}
\!\!\Pr\left({x_1}<x_m<\sqrt{\eta}{x_1} \right) = \!\!
\int \limits_{0}^{+\infty}f_{x_1}(x_1) \! \int \limits _{x_1}^{{\sqrt{\eta}}x_1} \!\! f_{x_m}(x_m)   \text{d}x_m \text{d}x_1 =\nonumber
\\
\!\! \int \limits_{0}^{+\infty} \!\!\!2\pi \lambda_s x e^{-\lambda_s \pi x^2}\! (e^{\!-\lambda_m \pi x^2} \!\!\!- e^{\!-\lambda_m \pi \eta x^2}\!) \;\!\! \text{d}x\! = \!\nonumber
\!\left(\frac{\lambda_s}{\lambda_s\! + \!\lambda_m} \!-\! \frac{\lambda_s}{\lambda_m \eta\! +\! \lambda_s} \right)\!\!. 
\end{array}
\end{equation}

Leveraging our previous calculations and exploiting the symmetry of the subcases, we establish the final expression for the probability of Case 5 as:
\begin{equation}
P_{\text{Case 5}} =2 \left( \frac{\lambda_s}{\lambda_s + \lambda_m} - \frac{\lambda_s}{\lambda_m \eta + \lambda_s} \right) - P_{\text{Case 4}}.
\end{equation}

\subsection{Case 6: Connection to the two SCells, both in UL and DL}
In this case, the UE associates with both SCells. 
%

The expressions for subcases 6.1 and 6.2 that define the subject case are given as follows:
\begin{equation}
	\begin{array}{c}
	\text{UL}: {P_d}\:\norm{X_1}^{-\alpha} > {P_d}\:\norm{X_2}^{-\alpha} > {P_d}\:\norm{X_m}^{-\alpha} \\  \text{DL}: {P_s}\:\norm{X_1}^{-\alpha} > \  {P_s}\:\norm{X_2}^{-\alpha}  > \  {P_m}\:\norm{X_m}^{-\alpha}
	\end{array}, \label{6.1}
\end{equation}
\begin{equation}
	\begin{array}{c}
	\text{UL}: {P_d}\:\norm{X_2}^{-\alpha} > {P_d}\:\norm{X_1}^{-\alpha} > {P_d}\:\norm{X_m}^{-\alpha} \\  \text{DL}: {P_s}\:\norm{X_2}^{-\alpha} > \  {P_s}\:\norm{X_1}^{-\alpha}  > \  {P_m}\:\norm{X_m}^{-\alpha}
	\end{array}. \label{6.2}
\end{equation}
	
Rearranging the above, we produce the following restrictive conditions (based on the DL) for both subcases:
\begin{equation}
\text{Subcase 6.1:}\ {x_1}<{x_2}<\left( \frac{P_s}{P_m}\right)^{1/\alpha}{x_m},
\end{equation}
\begin{equation}
\text{Subcase 6.2:}\ {x_2}<{x_1}<\left( \frac{P_s}{P_m}\right)^{1/\alpha}{x_m}.
\end{equation}

Focusing on the first subcase, we may continue with:
\begin{equation}
\begin{array}{l}
\Pr\left( { \sqrt{\eta}}{x_1} <  { \sqrt{\eta}}{x_2} <{x_m}\right) =
\int \limits _{0}^{+\infty} f_{x_1}(x_1) \!  \! \int \limits _{x_1}^{+\infty} f_{x_2}(x_2)  \\
 \int \limits _{ { \sqrt{\eta}}{x_2} } ^{+\infty} f_{x_m}(x_m)
 \text{d}x_m  \text{d}x_2  \text{d}x_1=
  \frac{\lambda_s}{\lambda_m{\eta}+\lambda_s}\frac{\lambda_s}{\lambda_m{\eta}+2\lambda_s}.
\end{array}
\end{equation}

Finally, we derive the sought probability for Case 6 as:
\begin{equation}
P_{\text{Case 6}} = \frac{2\lambda_s^2}{(\lambda_m{\eta}+\lambda_s)(\lambda_m{\eta}+2\lambda_s)}.
\end{equation}

\vspace{-10pt}
\section{Uplink Capacity Derivation} \label{CapDerivation}
Following the analysis outlined in the previous section, in the Dual-Connectivity scenario a user will attach to the two serving cells with respect to the cases summarized in Table~\ref{table1}. Since our study primarily focuses on evaluating the benefits of flexible user association schemes, the subsequent capacity expressions are derived specifically for a decoupled scenario and therefore concern Cases 3, 4, and 5. To characterize the gains made available by offering this new level of flexibility, the capacity of the decoupled link is compared to that for the association based on a downlink received power policy.

In the remainder, we begin with providing a general expression for the link capacity. Then, in order to average the capacity across the considered three cases, we employ the conditional distributions of distances to the closest SCell, the second closest SCell, and the MCell. The corresponding calculations are given in Appendix for all the three cases separately.
	%
	%
	
Clearly, the UL signal received in the cell $v$ may be expressed as $S_v = P_{tx} h_v \norm{X_v}^{-\alpha}$. Given the power splitting in the UL, the interference perceived from all the other users in the target scenario is expressed as $\sum_{i=1}^{n}\ P_d/2 \cdot h_i \cdot \norm{X_i}^{-\alpha}$, where $\norm{X_i}$ is the distance between the destination and the interferer $i$.

Further, the signal-to-interference ratio (SIR) is delivered by:
\begin{equation}
\text{SIR}_{\text{UL}} = \frac{P_d/2 \cdot h_v \cdot \norm{X_v}^{-\alpha}}{\sum_{i=1}^{n}\ P_d/2 \cdot h_i \cdot \norm{X_i}^{-\alpha} },
\end{equation}
where 
the interference is modeled based on the assumptions outlined in Section~\ref{system}. The UL throughput can then be characterized as:
\begin{equation}
\begin{array}{l}
C_{\text{UL}} = E_h\left [{B}\log_2(1 + \text{SIR}_{\text{UL}} ) \right] =\\
{B}\frac{B}{\ln2}\int \limits_{0}^{+\infty} \Pr(\ln(1 + \text{SIR}_{\text{UL}} )>t )\text{d}t =
\\
\frac{B}{\ln2}\int \limits_{0}^{+\infty} \Pr(h_v > {(e^t -1) x_v^\alpha I_x}) \text{d}t,
\end{array} \label{Capacity}
\end{equation}
where $I_x = \sum_{i=1}^{n} h_i \cdot \norm{X_i}^{-\alpha} $.
The expression for $C_{\text{UL}}$ in \eqref{Capacity} is derived by applying the following property: for $T>0$, $\mathbb{E}[T]=\int\limits_0^\infty tf(t)\mathrm{d}{t}=\int\limits_0^\infty (1-F(t))\mathrm{d}{t}$ (the proof follows from the integration by parts).

The total aggregate interference $I_x$ is calculated by using the Laplace transform and following the assumptions in Section~\ref{system} as well as the approach in \cite{UL}, but assuming that all the interfering users are maintaining the Dual-Connectivity. The reader is referred to \cite{UL} for the details of the proof in this interference derivation. The final expression for the interference component is thus:
\begin{equation}
\begin{array}{l}
\Pr(h_v > {(e^t -1) x_v^\alpha I_x}) =\\
\int \limits_{0}^{+\infty} e^{- \pi \lambda_{I_d} (e^t-1)^\frac{2}{\alpha}x^2\int \limits _{0}^{+\infty} \frac{\text{d}v}{1+v^\frac{\alpha}{2}} }f_x(x) \text{d}x,
\end{array}
\end{equation}
where $\lambda_{I_d} = p\lambda_{d}$ is the intensity of a thinned PPP for the interfering users, and $f_x(x)$ is the distribution of distances to the receiver. We emphasize that $f_x(x)$ is conditioned on the fact that the UE is located within the considered region (case) as well as it depends on the receiver type (SCell~1, SCell~2, or MCell). The probability $p$ is defined as the ratio between the number of interfering cells and the total number of users. Note that the dependencies on the OLPC parameter $\gamma$ have been disregarded here, since this study does not consider fractional path-loss compensation.

Further, the total throughput of one user is the aggregation of its throughput values over both links. Hence, our approach is to characterize each link individually by following a procedure similar to that in \cite{CapDUDe}. Based on the equation \eqref{regioscell} for Case 3, the corresponding UL throughput of the first connection (the closest SCell) is derived as:
	\begin{equation}
	\begin{array}{l}
	C_{\text{SCell 1} \vert \text{DUDe}} = \frac{B}{\ln 2}
	\bigintsss \limits_{0}^{+\infty} \bigintsss \limits_{0}^{+\infty} e^{- \pi \lambda_{I_d} (e^t-1)^\frac{2}{\alpha}x^2\int \limits _{0}^{+\infty} \frac{\text{d}v}{1+v^\frac{\alpha}{2}} } \times
	\\
	\frac{4\pi\lambda_s x e^{-\pi\lambda_sx^2}}{P_{\text{Case 3}}}
	\left( \frac{\lambda_me^{-\pi (\lambda_m \eta+\lambda_s){z^2} }}{\lambda_m + \lambda_s/\eta} \!\!- \!\! \frac{\lambda_me^{-\pi (\lambda_s+\lambda_m)\eta {z^2} }}{\lambda_m + \lambda_s} \right) \text{d}x \text{d}t,
	\label{C1DC}
	\end{array}
	\end{equation}
	while the second connection (the second closest SCell, (\ref{eqn:dist_SC2_case3})) gives us:
	\begin{equation}
	\begin{array}{l}
	C_{\text{SCell 2} \vert \text{DUDe}} = \frac{B}{\ln 2}
	\bigintsss \limits_{0}^{+\infty} \bigintsss \limits_{0}^{+\infty} e^{- \pi \lambda_{I_d} (e^t-1)^\frac{2}{\alpha}x^2\int \limits _{0}^{+\infty} \frac{\text{d}v}{1+v^\frac{\alpha}{2}} } \times
	\\
	\frac{4\pi\lambda_s x e^{-\pi\lambda_sx^2}}{P_{\text{Case 3}}}
	\left(e^{-\pi\lambda_m x_2^2} - e^{-\pi\lambda_m \eta x_2^2} -\right.
	\\ \left.
	\frac{\lambda_m \eta}{\lambda_s + \lambda_m \eta} e^{-\pi(\lambda_s/\eta+\lambda_m) x_2^2} +
	\frac{\lambda_m \eta}{\lambda_s + \lambda_m \eta} e^{-\pi(\lambda_s+\lambda_m \eta )x_2^2} \right)
	\text{d}x \text{d}t.
	\end{array}
	\end{equation}

Following the same procedure and utilizing the distance distribution derived in \eqref{regiomcell}, the capacity of the sub-optimal associated link is:
\begin{equation}
\begin{array}{l}
C_{\text{MCell} \vert \text{DRP}} = \frac{B}{\ln 2}
\bigintsss \limits_{0}^{+\infty} \bigintsss \limits_{0}^{+\infty} e^{- \pi \lambda_{I_d} (e^t-1)^\frac{2}{\alpha}x^2\int \limits_{0}^{+\infty} \frac{\text{d}v}{1+v^\frac{\alpha}{2}}} \times
\\
\frac{4\pi\lambda_mx e^{-\pi\lambda_mx^2}}{P_{\text{Case 3}}} \left( e^{-\pi \lambda_s{x^2}/\eta } - e^{-\pi \lambda_s{x^2} }\right) \times
\\
\left( 1-e^{-\pi \lambda_s{x^2}/\eta } \right) \text{d}x \text{d}t.
\label{cscell}
\end{array}
\end{equation}

The expression $\!\int \limits_{0}^{+\infty} \!\!\frac{\text{d}v}{1+v^\frac{\alpha}{2}}$ can be simplified for $\alpha >2$ as:
\begin{equation}
\int \limits_{0}^{+\infty} \frac{\text{d}v}{1+v^\frac{\alpha}{2}} = \frac{2\pi}{\alpha \sin \left(\frac{2\pi}{\alpha}\right)}.
\end{equation}
We note that the derivation of other decoupling cases may be easily driven by following the same procedure, with the only difference in the corresponding distance distributions provided in Appendix. It is therefore omitted here.


\section{Characteristic Performance Evaluation}

\subsection{Evaluation Conditions}

In the remainder of this paper, the typical performance of decoupled associations in the Dual-Connectivity setup is evaluated by representing the metrics that were derived analytically in the previous sections, as well as comparing them against those for various state-of-the-art alternatives.

The following counterpart baseline solutions have been considered by this study:
\begin{enumerate}
	\item \textit{Dual connectivity with no decoupled association:} this baseline is interesting to highlight to what extent the decoupled associations improve the overall performance of the UL Dual-Connectivity. Similar association methodologies have been utilized in other spectrum aggregation-specific works, as e.g., in \cite{6514952, IS-CA2}. This baseline is compared against Case 3 and Case 4 decoupling events, since Case 5 is a Dual-Connectivity association with inverse connections. In the following, this baseline is referred to as BL1.
	\item \textit{No spectrum aggregation (i.e., no Dual-Connectivity) and decoupled association:} this baseline is useful to justify the need for spectrum aggregation in the UL. In the past literature, some authors maintained that keeping more than one UL connection can be less power efficient, since UEs tend to be power limited \cite{R2-140054}. This baseline has been considered in the recent references \cite{DUDe, AsDude, PLDUDe} that indicated reasonable improvements for the UL transmissions. In the following, this baseline is referred to as BL2.
	\item \textit{UL carrier aggregation towards the strongest cell with no decoupled connections:} this baseline is dedicated to characterize the integration of the decoupled access with the spectrum aggregation techniques. Prior art in the field confirmed improvements in the UL carrier aggregation \cite{C1,J1} by enhancing the component carrier selection rules. In the following, this baseline is referred to as BL3.
\end{enumerate}

To analyze the improvements in the UL performance, spectral efficiency has been selected as the most representative parameter. The reason being is that the throughput calculations need to additionally consider the load situations in both the MCell and the SCells. If the loading is accounted for, then the HetNet-specific load balancing mechanisms can intervene with the improvements in signal quality brought by the decoupled associations.

For convenience, Table~\ref{Notation} summarizes the notation used throughout the mathematical component of this paper together with the actual values employed for the purposes of this performance evaluation.
\begin{table}
\caption{Our employed notation} \label{Notation}
	\centering
	\begin{tabular}{c l c}
		\toprule
		\textbf{Notation}             & \multicolumn{1}{c}{\textbf{Definition}} &\textbf{Value}
		\\ \midrule
		$\lambda_m$                    & Intensity of the MCells PPP        &                  $1.47\cdot10^{-5}$    \\
		$\lambda_s$                    & Intensity of the SCells PPP         &                  1-10                            \\
		$\lambda_d$                     & Intensity of the UEs PPP        &                  $0.037$                           \\
		$P_s$                          & SCell transmit power        &                 30\,dBm                               \\
		$P_m$                          & MCell transmit power                                                      &  43\,dBm  \\
		$P_d$                          & UE transmit power            &                 23\,dBm                                \\
		$\gamma$                        & Path-loss compensation factor    &          0                         \\
		$P_{\text{0}}$                  & OLPC parameter           &            23\,dBm                          \\
		$\alpha$                        & Path-loss exponent              &                  4                        \\
		$B$                          & System bandwidth& 20\,MHz\\
		Layout	& Grid side size & 1650\,m
		\\ \bottomrule
	\end{tabular}
	
\end{table}

\subsection{Numerical Results}
\begin{figure}
	\centering
	\includegraphics[width=\columnwidth]{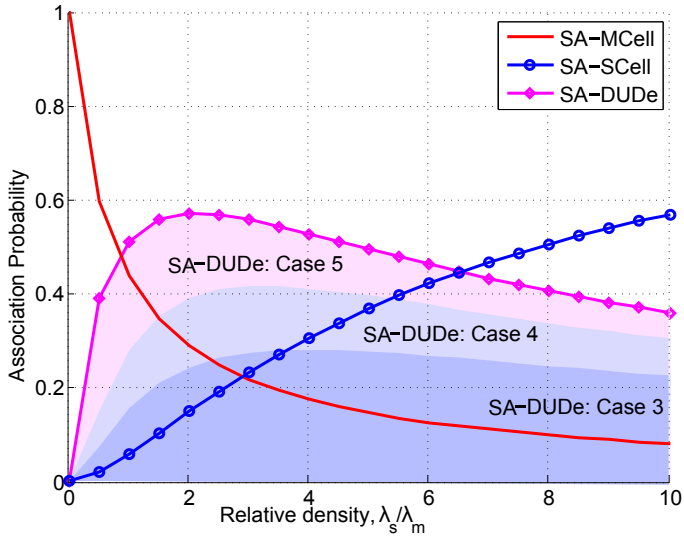}
	\caption{Association probability for Dual-Connectivity setup}
	\label{figure 5.1}
\end{figure}

First, the association probability is evaluated. To this end, Fig.~\ref{figure 5.1} demonstrates the probability of the cases shown in Table~\ref{table1}. Since the probabilities for every subcase are equal, the joint per-case probability is represented in the plots. Also, since Cases 3, 4, and 5 all include the events related to the decoupled associations, we have collected them under a single curve, named DUDe (Downlink and Uplink Decoupled). Cases 1 and 2 that correspond to the classical Dual-Connectivity association situations are aggregated as the DualConn curve. Finally, spectrum aggregation across the two SCells (Case 6) is illustrated with the SCell curve in the figure.

In the Dual-Connectivity, the probability that the decoupled events occur cannot be neglected. In fact, decoupled associations are almost 40\% more probable than the MCell associations (Cases 1 and 2, Dual-Connectivity) and over 43\% to 22\%  more probable than the SCell coupled associations (Case 6) for low numbers of SCells ($\lambda_s/\lambda_m=2$ and $\lambda_s/\lambda_m=4$, respectively). Generally, as the number of SCells increases, the probability of having coupled associations to the SCell grows. Based on this observation, there is a considerably high chance that the legacy association policies in the Dual-Connectivity become sub-optimal, since no flexibility is allowed.

\begin{figure}
	\centering
	\includegraphics[width=\columnwidth]{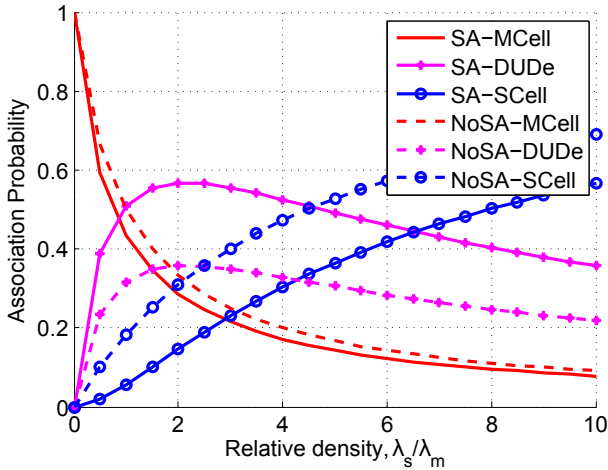}
	\caption{Association probabilities for decoupled Dual-Connectivity vs. single connectivity.}
	\label{figure 5.2}
\end{figure}

As the association regions depend largely on the user proximity to the serving cell and the cell transmit power, contrasting the connection probabilities for the Dual-Connectivity against the Carrier Aggregation baseline is identical to comparing it with the association on a single carrier with one cell. Accordingly, Fig.~\ref{figure 5.2} compares the association probability for the Dual-Connectivity with the classical single cell association case that allows for decoupled connections (corresponds to BL2 as per our above description).

In the Dual-Connectivity, the number of association possibilities increases as compared to the single cell attachment, and the probability region is more spread. For the sake of clarity, in this figure the represented situations are also collected together under the Dual-Connectivity classical association setup, DUDe, and the SCell associations (Case 6). Since multi-connectivity increases the number of combinations to be considered, decoupled associations are more probable than in the single connectivity case. Said increase in decoupled access comes along with the reduction in 'pure' SCell associations (Case 6) or classical Dual-Connectivity configurations: MCell and SCell spectrum aggregation.

\begin{figure}
	\centering
	\includegraphics[width=\columnwidth]{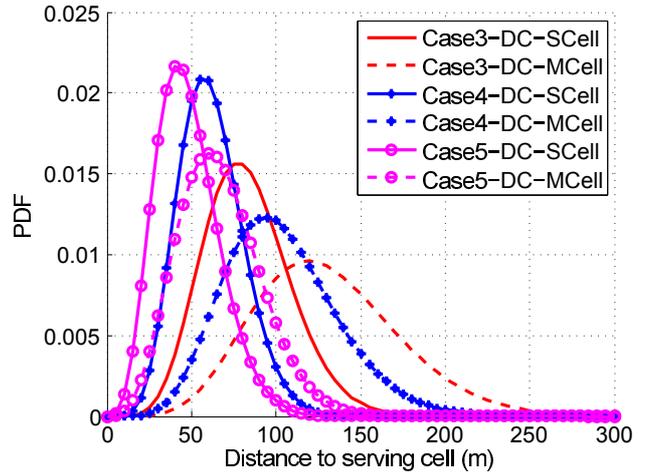}
	\caption{Distance distribution for all decoupled regions}
	\label{figure 5}
\end{figure}

\begin{figure}
	\subfigure[UL spectral efficiency for decoupled link only]{\label{fig:Cap1}\includegraphics[width=\columnwidth]{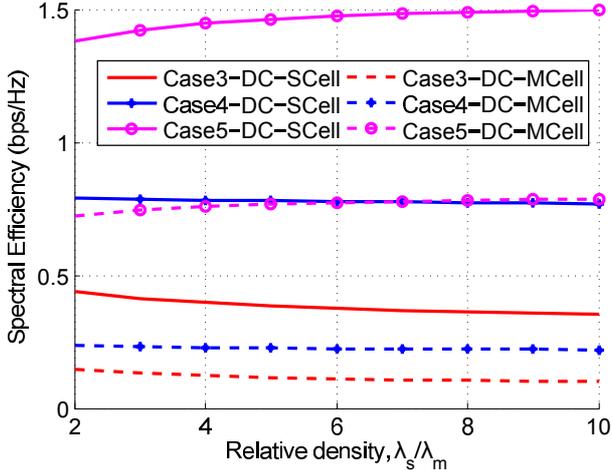}}\\
	\subfigure[UL spectral efficiency gain for DUDe Dual-Connectivity $\lambda_s/\lambda_m=5$]{\label{fig:Cap2}\includegraphics[width=\columnwidth]{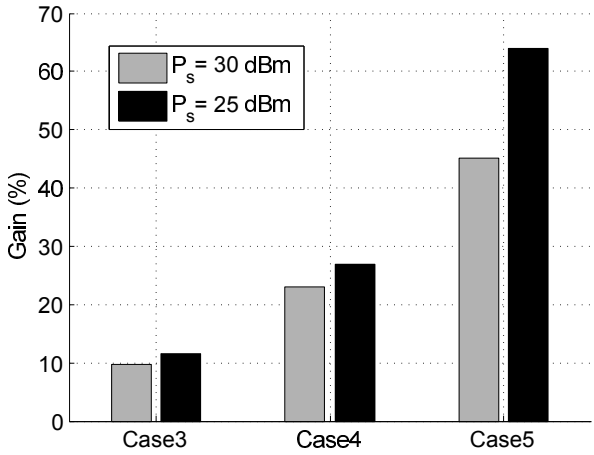}}
	\caption{UL capacity gain: decoupled association vs. sub-optimal association} \label{fig:Capacity1}
\end{figure}

The UL throughput gains depend largely on the distance distribution to the serving cell, shown in Fig.~\ref{figure 5}, where all the cases related to the decoupling events are represented. The distance distribution in question is conditioned on the association region, which corresponds to equations \eqref{regioscell} and \eqref{regiomcell} of our mathematical analysis in Section~\ref{CapDerivation}. For all these decoupling cases, it is confirmed that the distance to the SCell is much shorter than that to the MCell, thus resulting in the optimal UL association that allows the UE to maximize its spectral efficiency. Here, Case 3 typically has the longest distance to the SCell because the decoupled connection is performed on the second link -- to the second closest SCell -- and the first link is associated with the dominant SCell. Accordingly, Fig.~\ref{fig:Capacity1} reports on the UL spectral efficiency for one user (see equation \eqref{Capacity}); it compares the decoupled association in Cases 3, 4, and 5 with the sub-optimal association to the MCell.

In particular, Fig.~\ref{fig:Cap1} focuses specifically on the capacity of the decoupled link, while Fig.~\ref{fig:Cap2} highlights the gain obtained in the aggregate spectral efficiency, while accounting for both connections and different $P_s$ configurations. It is worth noting that even though the probability that such decoupling cases occur is very much dependent on the relative density of the cells, the resulting spectral efficiency values for each case maintain constant regardless of the number of SCells deployed in the scenario of interest. On the other hand, lower SCell transmit power does have an impact on the association regions, separating the UE from the MCell even further and thus leading to higher spectral efficiency gains, see Fig~\ref{fig:Cap2}.

It is evident that altering the association policy in the UL and hence allowing to decouple the second connection to the SCell enables the capacity boost as well as augments the gains of the Dual-Connectivity further. Such gains are more noticeable for Cases 4 and 5, where the association region demonstrates a drastic change in the distance distribution by having almost 40\,m of difference in the average distance to the serving cell, as shown in Fig.~\ref{fig:Cap1}. In particular, for $P_s=30$\,dBm the SIR is enhanced by more than 7\,dB in all three cases; these gains stem from an improvement in the UL received power, since the users are transmitting to a closer cell. In fact, when following the DL received power association policies, the UE signal quality in the decoupling regions can be very low (-3\,dB in Case 3, sub-optimal setup), thus resulting in a poorly configured UL connection.

\begin{figure}
	\centering
	\includegraphics[width=\columnwidth]{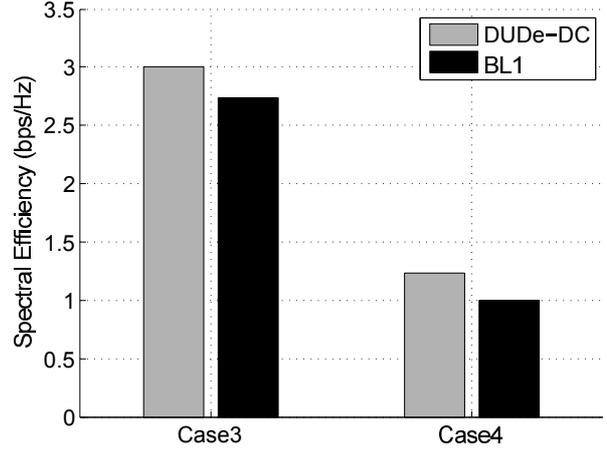}
	\caption{Comparing Dual-Connectivity with DUDe vs. conventional Dual-Connectivity; $\lambda_s/\lambda_m=5$}
	\label{DUDevsDC}
\end{figure}

A major advantage of the Dual-Connectivity is in the capacity increase that the user is experiencing. However, for users positioned in the decoupling regions this capacity may remain sub-optimal, since the received power is strongly attenuated by the distance to the serving cell. Along these lines, Fig.~\ref{DUDevsDC} compares two Dual-Connectivity schemes: one with the DUDe-enabled associations and another representing a classical Dual-Connectivity association (BL1). In this discussion, the decoupling Cases 3 and 4 are considered, while Case 5 is disregarded because the resultant serving cells are the same as for the conventional Dual-Connectivity scheme, but with both links inverted. In particular, Case 3 indicates the largest improvement, as corresponds to the association region where the user is closer to the SCell. Our results demonstrate benefits in the spectral efficiency as compared to the classical Dual-Connectivity approach, with a more than 0.7\,bps/Hz increase.

\begin{figure}
	\centering
	\includegraphics[width=\columnwidth]{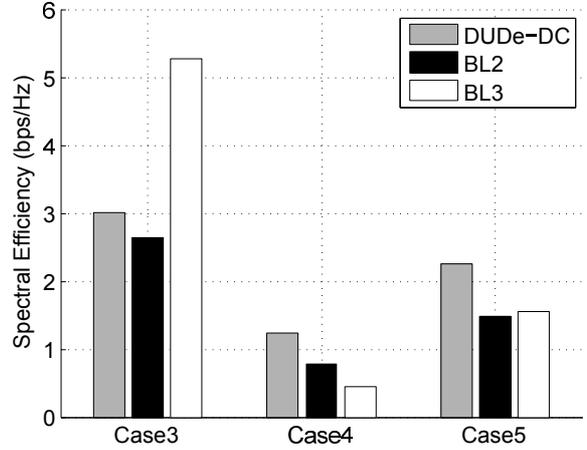}
	\caption{Comparison of Dual-Connectivity with DUDe vs. single connectivity with DUDe and Carrier Aggregation with DRP; $\lambda_s/\lambda_m=5$}
	\label{DUDevsSC}
\end{figure}

It is also useful to compare the performance of the Dual-Connectivity with that of the single connectivity, as well as assess the UL gains of maintaining more than one connection. In this case, the comparison is conducted with a single cell (but allowing for decoupled connections, BL2), thus enabling to maximize the UL performance. To this end, Fig.~\ref{DUDevsSC} presents the analysis of the UL throughput for the Dual-Connectivity and contrasts it against that for the decoupled association with a single cell. The gains for Case 3 are less noticeable, since the UE's first association is very strong in the given association region. Nevertheless, Fig.~\ref{DUDevsSC} justifies the gains of the Dual-Connectivity, as they are clearly visible when compared to the single best association case. It is also evident that even though the UE is forced to perform power splitting between the two cells, the resulting throughput boost is significant.

Finally, Fig.~\ref{DUDevsSC} compares the performance of the proposed setup against the well-known baseline for Carrier Aggregation, where association is done by following the DRP association rules (i.e., no decoupled associations, BL3). Given the coupled association policy, the baselines in Cases 4 and 5 have two UL connections to the MCell. Granted that the strongest DL connection in Case 3 is to the SCell~1, both UL connections are kept to the closest SCell. Apparently, if the UE aggregates spectrum with Carrier Aggregation to a sub-optimal cell (Cases 4 and 5), the capacity benefits remain limited, mainly because of the proximity to the serving cell. On the contrary, when decoupled associations are allowed, the UEs utilize spectrum aggregation more flexibly, which in turn allows to maximize the UL capacity.

It is also worth studying the behavior of Case 3 more closely, since it improves the spectral efficiency performance when aggregating carriers in one cell. This is because for this association region the decoupling event is in the second connection, thus resulting in highly sub-optimal split on both ULs. Intuitively, connecting all the links to the closest cell will result in the best performance, as the received power will always remain maximized. However, the use of the Dual-Connectivity has the potential to go one step beyond Carrier Aggregation and unlock benefits from the extra capacity of the neighboring cells. In this sense, Carrier Aggregation will always need two or more component carriers to perform spectrum aggregation, while in the Dual-Connectivity the aggregation can be performed within the same carrier.

\section{Conclusions}
This work has studied the advantages of allowing for decoupled associations in the Dual-Connectivity scenarios, where the users are enabled to simultaneously consume radio resources from two serving cells. Spectrum aggregation techniques have always been more applicable for the DL, while in the UL they showed limited benefits because most UEs are power constrained. Aiming to improve the UE throughput as well as the overall user connectivity experience, we proposed to decouple the UL connection and introduced the UL-specific association rules in the context of multi-connectivity for HetNets. Our results empower the user to experience the maximum flexibility when deciding which cells to aggregate resources from, as well as fully benefit from the Dual-Connectivity advantages in the UL.

The HetNet system in question was modeled analytically by employing stochastic geometry. Correspondingly, a set of two SCells and one MCell was considered and we demonstrated that the number of decoupled events is large, resulting in high probabilities to leverage one of these when selecting the serving cells. Overall, the main conclusions of this work can be summarized as follows:
\begin{enumerate}
	\item The Dual-Connectivity scenarios display a total of twelve types of association regions that can be generally reduced to six cases, three of which relate to the decoupled association. We have comprehensively studied each of these identified association regions and derived convenient closed-form expressions for the probabilities in the setup with three different cells forming a cluster. Our considered association policy was the maximum received power per link. We have observed that for the three cases that involve decoupled events, their occurrence probability increases with the growing SCell density, almost 40\% with respect to the conventional Dual-Connectivity association cases.
	\item We have derived the distributions of distance to all the three cells conditioned on each of the decoupled association cases, which indicated a clear reduction in the user path-loss when connecting to a SCell, thus bringing along higher received powers.
	\item We have studied the UL spectral efficiency gains with multiple connections to two different serving cells by comparing the outcomes of our mathematical analysis with several alternative solutions: the DL received power based association rule in the Dual-Connectivity, the single connectivity with a decoupled association, and Carrier Aggregation to the MCell. We concluded that the best form of spectrum aggregation for users in the decoupled regions is to allow for splitting the UL and the DL, since significant performance gains can be made available then.
\end{enumerate}

\vspace{-15pt}
\section*{Acknowledgment}
This work was supported by Ericsson 5G and Tactile Internet industry grant to King's College London as well as by the ICT-SOLDER project, www.ict-solder.eu, FP7 project number 619687. The work of O. Galinina was supported with a personal research grant by the Finnish Cultural Foundation. The work of S. Andreev was supported in part by a Postdoctoral Researcher grant from the Academy of Finland and in part by a Jorma Ollila grant from Nokia Foundation.

\vspace{-15pt}
\section*{Appendix}
In this Appendix, we collect the derivations of the conditional distributions for the Dual-Connectivity regions, including Case 3, Case 4, and Case 5, respectively. Each set distributions includes an expression for the distances to the MCell as well as to the closest and the second closest SCell (except for Case 5, which assumes no connection to the second SCell).

\vspace{-10pt}
\subsection{Case 3}
The region for Case 3 is defined by the expression $ {x_1}<\frac{x_m}{ \sqrt{\eta}}< {x_2}< {x_m}$, where $x_1$ corresponds to the closest SCell. Therefore, the CDF of distances $x_1$ may be expressed as:
	\begin{equation}
	\begin{array}{l}
	F_{x_{s1\vert \text{Case 3}} }(x)=\Pr \left(x_1< x \left \vert {x_1}<\frac{x_m}{ \sqrt{\eta}}< {x_2}< {x_m} \right. \right) =
	\end{array} \nonumber
	\end{equation}
	\begin{equation}
	\begin{array}{l}
	\frac{2}{P_{\text{Case 3}}}
	\int \limits_{0}^{x} f_{x_1}(x_1) \int \limits _{x_1\sqrt{\eta}}^{\infty} f_{x_m}(x_m)
	\int \limits_{\frac{x_m}{\sqrt{\eta}}}^{x_m} f_{x_2}(x_2) \text{d}x_2 \text{d}x_m \text{d}x_1=
	\end{array} \nonumber
	\end{equation}
	\begin{equation}
	\begin{array}{l}
	\frac{2}{P_{\text{Case 3}}}
	\int \limits^{x}_{0} \!\!
	\left( \frac{\lambda_me^{-\pi (\lambda_m \eta+\lambda_s){z^2} }}{\lambda_m + \lambda_s/\eta} \!\!- \!\! \frac{\lambda_me^{-\pi (\lambda_s+\lambda_m)\eta {z^2} }}{\lambda_m + \lambda_s} \right) \! \! f_{x_1}\!(z)\text{d}z.
	\end{array} \nonumber
	\end{equation}
	
	Differentiating the CDF, we may obtain the pdf of the distances to the serving cell conditioned on the events of subcase 3.1 as follows:
	\begin{equation}
	\begin{array}{l}
	f_{x_{s1\vert \text{Case 3}} }\!(x)\!=\!
	\!\frac{2}{P_{\text{Case 3}}}\!
	\left( \! \frac{\lambda_me^{-\pi (\lambda_m+\lambda_s/ \eta){x^2} }}{\lambda_m + \lambda_s/\eta} \!\!- \!\! \frac{\lambda_me^{-\pi (\lambda_s+\lambda_m) {x^2} }}{\lambda_m + \lambda_s} \right) \! \! f_{x_1}\!(x),
	\end{array} \label{regioscell}
	\end{equation}
	where $f_{x_1}\!(x)=2\pi\lambda_sx e^{-\pi\lambda_s x^2}$.
		
In turn, for the distances $x_2$ to the second closest SCell -- by using the same condition -- we may establish the CDF as:
\textcolor{black}{\begin{equation}
	\begin{array}{l}
	F_{x_{s2\vert \text{Case 3}} }(x)=\Pr \left(x_2< x \left \vert {x_1}<\frac{x_m}{ \sqrt{\eta}}< {x_2}< {x_m} \right. \right) =\\
	\frac{2}{P_{\text{Case 3}}}
	\int \limits_{0}^{x} f_{x_2}(x_2) \int \limits _{x_2}^{x_2\sqrt{\eta}} f_{x_m}(x_m)
	\int \limits_{0}^{\frac{x_m}{\sqrt{\eta}}} f_{x_1}(x_1) \text{d}x_1 \text{d}x_m \text{d}x_2=
	\\
	\frac{2}{P_{\text{Case 3}}}
	\int \limits^{x}_{0} \!\!
	\left(e^{-\pi\lambda_m x_2^2} - e^{-\pi\lambda_m \eta x_2^2}
	- \frac{\lambda_m \eta}{\lambda_s + \lambda_m \eta} e^{-\pi(\lambda_s/\eta+\lambda_m) x_2^2} +\right.
	\end{array} \nonumber
	\end{equation}
	\begin{equation}
	\begin{array}{l}
	\left.
	\frac{\lambda_m \eta}{\lambda_s + \lambda_m \eta} e^{-\pi(\lambda_s+\lambda_m \eta )x_2^2}\right)
	f_{x_2}(x_2) \text{d}x_2,
	\end{array} \nonumber
	\end{equation}
where the corresponding expression for the pdf is given by:
	\begin{equation}
	\begin{array}{l}
	f_{x_{s2\vert \text{Case 3}} }\!(x)\!=\!
	\!\frac{2}{P_{\text{Case 3}}}\!
	\left(e^{-\pi\lambda_m x_2^2} - e^{-\pi\lambda_m \eta x_2^2} -\right.
	\\ \left.
	\frac{\lambda_m \eta}{\lambda_s + \lambda_m \eta} e^{-\pi(\lambda_s/\eta+\lambda_m) x_2^2} +
	\frac{\lambda_m \eta}{\lambda_s + \lambda_m \eta} e^{-\pi(\lambda_s+\lambda_m \eta )x_2^2} \right) f_{x_2}(x_2), \label{eqn:dist_SC2_case3}
	\end{array}
	\end{equation}
	where $f_{x_2}\!(x)=2\pi\lambda_sx e^{-\pi\lambda_s x^2}$.}

\textcolor{black}{Further, the CDF of the distances to the serving cell for the sub-optimal association option -- being the MCell in this case -- may be expressed as:
	\begin{equation}
	\begin{array}{l}
	F_{x_{m\vert \text{Case 3}} }(x)=\Pr \left(x_m< x \left \vert {x_1}<\frac{x_m}{ \sqrt{\eta}}< {x_2}< {x_m} \right. \right) =\\
	\frac{2}{P_{\text{Case 3}}}
	\int \limits_{0}^{x} f_{x_m}(x_m) \int \limits _{0}^{\frac{x_m}{ \sqrt{\eta}}} f_{x_1}(x_1)
	\int \limits_{\frac{x_m}{\sqrt{\eta}}}^{x_m} f_{x_2}(x_2) \text{d}x_2 \text{d}x_1 \text{d}x_m=\\
	\frac{2}{P_{\text{Case 3}}}
	\int \limits^{x}_{0} \!\!
	\left( e^{-\pi \lambda_s{x_m^2}/\eta } \!-\!e^{-\pi \lambda_s{x_m^2} }\right)
	\!\!\left( 1\!-\!e^{-\pi \lambda_s{x_m^2}/\eta } \right)\!\!f_{x_m}\!(x_m)\text{d}x_m,
	\end{array} \nonumber
	\end{equation}
	where the pdf may be obtained as follows:
	%
	\begin{equation}
	\begin{array}{r}
	f_{x_{m\vert \text{Case 3}} }(x)=\frac{2}{P_{\text{Case 3}}}
	\left( e^{-\pi \lambda_s{x^2}/\eta } - e^{-\pi \lambda_s{x^2} }\right) \times
	\\
	\left( 1-e^{-\pi \lambda_s{x^2}/\eta } \right)f_{x_m}(x),
	\end{array} \label{regiomcell}
	\end{equation}
	where $f_{x_m}\!(x)=2\pi\lambda_mx e^{-\pi\lambda_m x^2}$.
}

\vspace{-15pt}
\subsection{Case 4}
For brevity, we will further refer to $f_{x_1}\!(x)$, $f_{x_2}\!(x)$, and $f_{x_m}\!(x)$ as indicated in the previous subsection. We note that Case 4 is defined by the set of inequalities $ {x_1}<{x_2}< {x_m}< \sqrt{\eta} {x_1}$. Hence, we may write down the expression for the CDF of the distances $x_1$ to the closest SCell as follows:
	\begin{equation}
	\begin{array}{l}
	F_{x_{s1\vert \text{Case 4}} }(x)=\Pr \left(x_1< x \left \vert {x_1}<{x_2}< {x_m}< \sqrt{\eta} {x_1} \right. \right) =
	\\
	\frac{2}{P_{\text{Case 4}}}
	\int \limits_{0}^{x} f_{x_1}(x_1) \int \limits _{x_1}^{x_1\sqrt{\eta}} f_{x_2}(x_2)
	\int \limits_{x_2}^{x_1\sqrt{\eta}} f_{x_m}(x_m) \text{d}x_m \text{d}x_2 \text{d}x_1=
	\\
	\frac{2}{P_{\text{Case 4}}}
	\int \limits_{0}^{x} f_{x_1}(x_1)
	(\frac{\lambda_s}{\lambda_m+\lambda_s}e^{-\pi(\lambda_m +\lambda_s)x_1^2}-
	\frac{\lambda_s}{\lambda_m+\lambda_s}
	\\
	e^{-\pi(\lambda_m +\lambda_s)\eta x_1^2}-e^{-\pi(\lambda_m \eta+\lambda_s )x_1^2} +
    \\
    e^{-\pi(\lambda_m \eta+\lambda_s \eta) x_1^2}) \text{d}x_1,
	\end{array} \nonumber
	\end{equation}
	as well as produce the corresponding pdf, which may be obtained by differentiating:
	\begin{equation}
	\begin{array}{l}
	f_{x_{s1\vert \text{Case 4}} }\!(x)\!=
	\frac{2}{P_{\text{Case 4}}}
	f_{x_1}\!(x) \left(\frac{\lambda_s}{\lambda_m+\lambda_s}e^{-\pi(\lambda_m +\lambda_s)x_1^2}-\right.
	\\
	\left. \frac{\lambda_s}{\lambda_m+\lambda_s}e^{-\pi(\lambda_m +\lambda_s) \eta x_1^2}-
	e^{-\pi(\lambda_m \eta+\lambda_s )x_1^2} +e^{-\pi(\lambda_m \eta+\lambda_s \eta) x_1^2}\right) .
	\end{array}
	\end{equation}

	The CDF of the distances to the second closest SCell is given by:
\textcolor{black}{
	\begin{equation}
	\begin{array}{l}
	F_{x_{s2\vert \text{Case 4}} }(x)=\Pr \left(x_2< x \left \vert {x_1}<{x_2}< {x_m}< \sqrt{\eta} {x_1} \right. \right) =
	\\
	\frac{2}{P_{\text{Case 4}}}
	\int \limits_{0}^{x} f_{x_2}(x_2) \int \limits _{\frac{x_2}{\sqrt{\eta}}}^{x_2} f_{x_1}(x_1)
	\int \limits_{x_2}^{x_1\sqrt{\eta}} f_{x_m}(x_m) \text{d}x_m \text{d}x_1\text{d}x_2=
	\end{array} \nonumber
	\end{equation}
	\begin{equation}
	\begin{array}{l}
	\frac{2}{P_{\text{Case 4}}}f_{x_2}(x_2)
	\left(e^{-\pi(\lambda_s/\eta+\lambda_m) x_2^2}
	- e^{-\pi(\lambda_m+\lambda_s)x_2^2} \right.
	\\ \left.
	+ \frac{\lambda_s}{\lambda_s + \lambda_m\eta} e^{-\pi(\lambda_s+\lambda_m\eta) x_2^2}
	- \frac{\lambda_s}{\lambda_s + \lambda_m\eta} e^{-\pi(\lambda_s/\eta+\lambda_m)x_2^2} \right)
	\text{d}x_2.
	\end{array} \nonumber
	\end{equation}
	and, hence, the pdf of the distances $x_2$ is:
	\begin{equation}
	\begin{array}{l}
	f_{x_{s2\vert \text{Case 4}} }\!(x)\!=
	\frac{2}{P_{\text{Case 4}}}
	f_{x_2}(x_2)
	\left(e^{-\pi(\lambda_s/\eta+\lambda_m) x_2^2}
	- e^{-\pi(\lambda_m+\lambda_s)x_2^2} \right.
	\\ \left.
	+ \frac{\lambda_s}{\lambda_s + \lambda_m\eta} e^{-\pi(\lambda_s+\lambda_m\eta) x_2^2}
	- \frac{\lambda_s}{\lambda_s + \lambda_m\eta} e^{-\pi(\lambda_s/\eta+\lambda_m)x_2^2} \right).
	\end{array}
	\end{equation}}

Using the same conditions of Case 4, we may obtain the following for $x_m$:
\textcolor{black}{\begin{equation}
\begin{array}{l}
F_{x_{m\vert \text{Case 4}} }(x)=\Pr \left(x_m< x  \left \vert {x_1}<{x_2}< {x_m}< \sqrt{\eta} {x_1} \right. \right)= 
\end{array} \nonumber
	\end{equation}
	\begin{equation}
	\begin{array}{l}
\frac{2}{P_{\text{Case 4}}}
\int \limits_{0}^{x}  f_{x_m}(x_m)  \int \limits _{\frac{x_m}{ \sqrt{\eta} }}^{x_m}  f_{x_1}(x_1)
\int \limits_{x_1}^{x_m}  f_{x_2}(x_2)   \text{d}x_2  \text{d}x_1 \text{d}x_m=
\end{array} \nonumber
	\end{equation}
	\begin{equation}
	\begin{array}{l}
\frac{2}{P_{\text{Case 4}}}
\int \limits_{0}^{x}  f_{x_m}(x_m)
(\frac{1}{2}e^{-\pi 2\lambda_s{x_m^2}/\eta} +\frac{1}{2}e^{-\pi 2\lambda_s x_m^2}\\
-e^{-\pi \lambda_s(1+1/\eta) x_m^2})   \text{d}x_m,
 \end{array} \nonumber
\end{equation}}
and the pdf for the distances $x_m$ follows the expression:
\textcolor{black}{\begin{equation}
\begin{array}{r}
f_{x_{m\vert \text{Case 4}} }(x)=\frac{2}{P_{\text{Case 3}}}
(\frac{1}{2}e^{-\pi 2\lambda_s{x_m^2}/\eta} +\frac{1}{2}e^{-\pi 2\lambda_s x_m^2}
\\
-e^{-\pi \lambda_s(1+1/\eta) x_m^2}) f_{x_m}(x).
\end{array}
\end{equation}}
%
%
%

\vspace{-20pt}
\subsection{Case 5}
Since Case 5 does not involve any associations to the second SCell, we provide our below derivations only for the distances $x_1$ and $x_m$ that determine the connections to the closest SCell and to the MCell, respectively. Following the same logic as above, we write down the expression for the CDF of the distances $x_1$ as:
\textcolor{black}{
	\begin{equation}
	\begin{array}{l}
	F_{x_{s\vert \text{Case 5}} }(x)=\Pr \left(x_1< x \left \vert {x_1}<x_m<{ \sqrt{\eta}}x_1,{x_2}> {x_m} \right. \right) =
	\end{array} \nonumber
	\end{equation}
	\begin{equation}
	\begin{array}{l}
	\frac{2}{P_{\text{Case 5}}}
	\int \limits_{0}^{x} f_{x_1}(x_1) \int \limits _{x_1}^{x_1\sqrt{\eta}} f_{x_m}(x_m)
	\int \limits_{x_m}^{\infty} f_{x_2}(x_2) \text{d}x_2 \text{d}x_m \text{d}x_1=
	\end{array} \nonumber
	\end{equation}
	\begin{equation}
	\begin{array}{l}
	\frac{2}{P_{\text{Case 5}}}
	\int \limits_{0}^{x} f_{x_1}(x_1) \frac{\lambda_m}{\lambda_s+\lambda_m}
	(e^{-\pi (\lambda_s+\lambda_m){x_1^2}} - e^{-\pi (\lambda_s+\lambda_m){x_1^2}\eta} ) \text{d}x_1,
	\end{array} \nonumber
	\end{equation}
	where the corresponding pdf may be derived as:
	\begin{equation}
	\begin{array}{l}
	f_{x_{s\vert \text{Case 5}} }\!(x)\!=\!
	\!\frac{2f_{x_1}\!(x)}{P_{\text{Case 5}}}\!
	\frac{\lambda_m}{\lambda_s+\lambda_m}
	(e^{-\pi (\lambda_s+\lambda_m){x_1^2}} - e^{-\pi (\lambda_s+\lambda_m){x_1^2}\eta} ) .
	\end{array}
	\end{equation}
	}
	
For the distances $x_m$, we establish the CDF in the form:
\textcolor{black}{\begin{equation}
	\begin{array}{l}
	F_{x_{m\vert \text{Case 5}} }(x)=\Pr \left(x_m< x \left \vert \frac{x_m}{ \sqrt{\eta}}<x_1<x_m,{x_2}> {x_m} \right. \right) =\\
	\!\frac{2}{P_{\text{Case 5}}}
	\int \limits_{0}^{x} f_{x_m}(x_m) \int \limits _{ \frac{x_m}{\sqrt{\eta}} }^{x_m} f_{x_1}(x_1)
	e^{-\pi \lambda_s{x_m^2}} \text{d}x_m=
	\end{array} \nonumber
	\end{equation}
	\begin{equation}
	\begin{array}{l}
	\!\frac{2}{P_{\text{Case 5}}}
	\int \limits_{0}^{x} f_{x_m}(x_m)
	e^{-\pi \lambda_s{x_m^2}}(e^{-\pi \lambda_s{x_m^2}/\eta}-e^{-\pi \lambda_s{x_m^2}}) \text{d}x_2 \text{d}x_1 \text{d}x_m,
	\end{array} \nonumber
	\end{equation}
and the respective pdf is thus given by:
	\begin{equation}
	\begin{array}{r}
	f_{x_{m\vert \text{Case 5}} }(x)=\frac{2}{P_{\text{Case 5}}}
	e^{-\pi \lambda_s{x_m^2}}(e^{-\pi \lambda_s{x_m^2}/\eta}-e^{-\pi \lambda_s{x_m^2}})f_{x_m}(x_m) .
	\end{array}
	\end{equation}
}
\vspace{-15pt}

\bibliographystyle{IEEEtran}
\bibliography{bibliography}
\vspace{-1.5cm}
\begin{IEEEbiography}[{\includegraphics[width=1in,height=1.25in,clip,keepaspectratio]{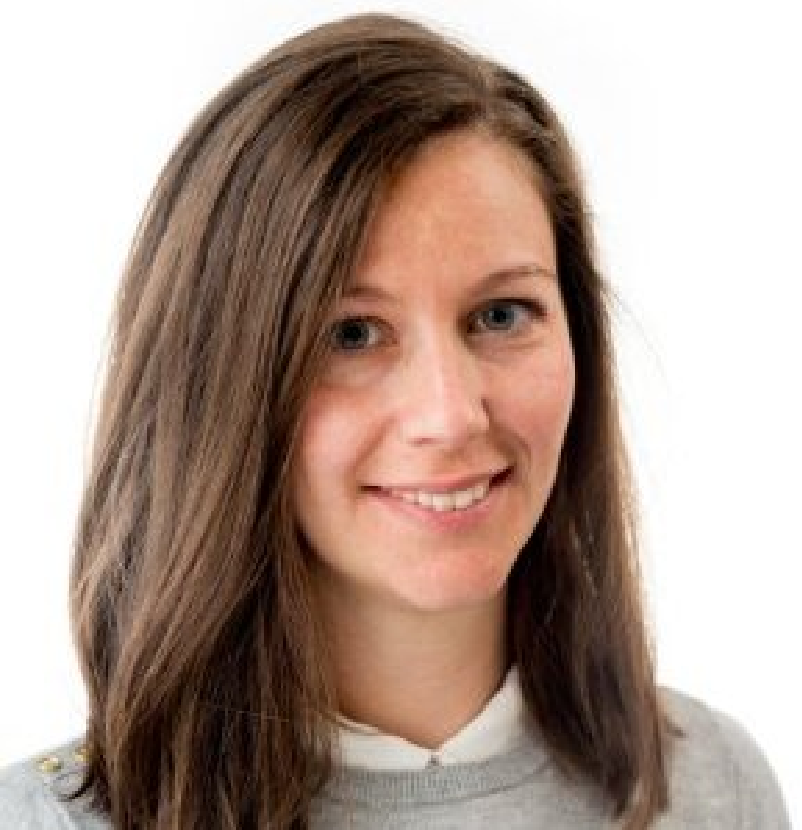}}]{Maria A. Lema}
		is currently a post-doctoral researcher at King’s College London, London, U.K., for the 5G Tactile Internet project together with Ericsson. She has been involved in the use cases definition for 5G, working together with the verticals and other telecom players to identify the main requirements and challenges to successfully bring 5G into other industries. She obtained her M.Sc. in Telecommunications Management (2010) and Ph.D in Wireless communications (2015) from Universitat Politecnica de Catalunya. Her major topic of research in her PhD was LTE-A system level studies, particularly uplink studies with carrier aggregation techniques. Nowadays, her research interests are 4G/5G Radio Access Network, Internet of Things and the Tactile Internet.
\end{IEEEbiography}
\vspace{-1.5cm}
\begin{IEEEbiography}[{\includegraphics[width=1in,height=1.25in,clip,keepaspectratio]{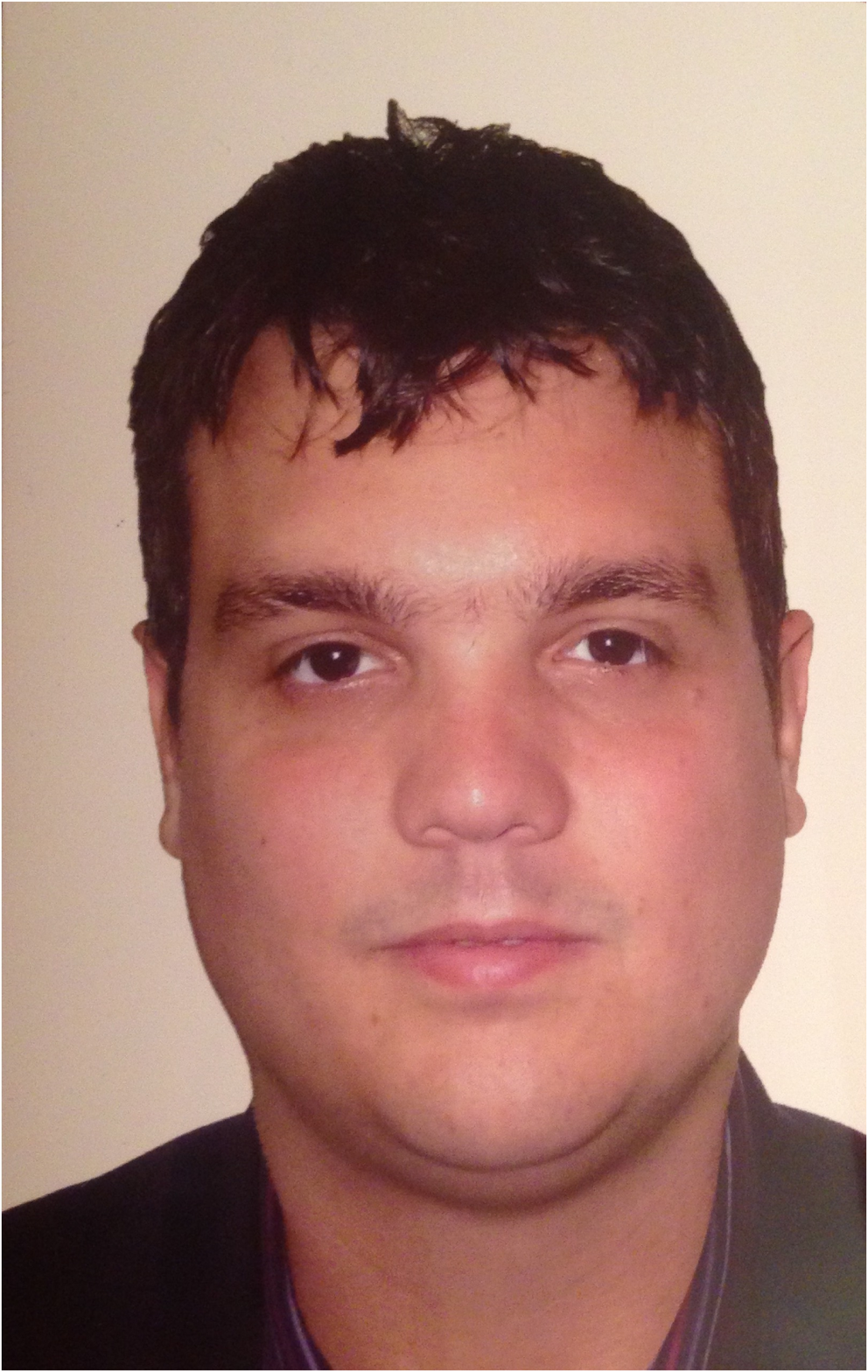}}]{Enric Pardo}
		is a Ph.D. candidate at the Centre for Telecommunications Research at King's College London, London, U.K.. He received is B.Sc. degree in Telecommunications Engineering from Universitat Politecnica de Catalunya (Spain) in 2016. He has recently started his Ph.D. studies in the area of multi-connectivity in 5G Networks. His area of study also covers mathematical analysis with the use of stochastic geometry for network performance evaluation.
\end{IEEEbiography}
\vspace{-1.5cm}
\begin{IEEEbiography}[{\includegraphics[width=1in,height=1.25in,clip,keepaspectratio]{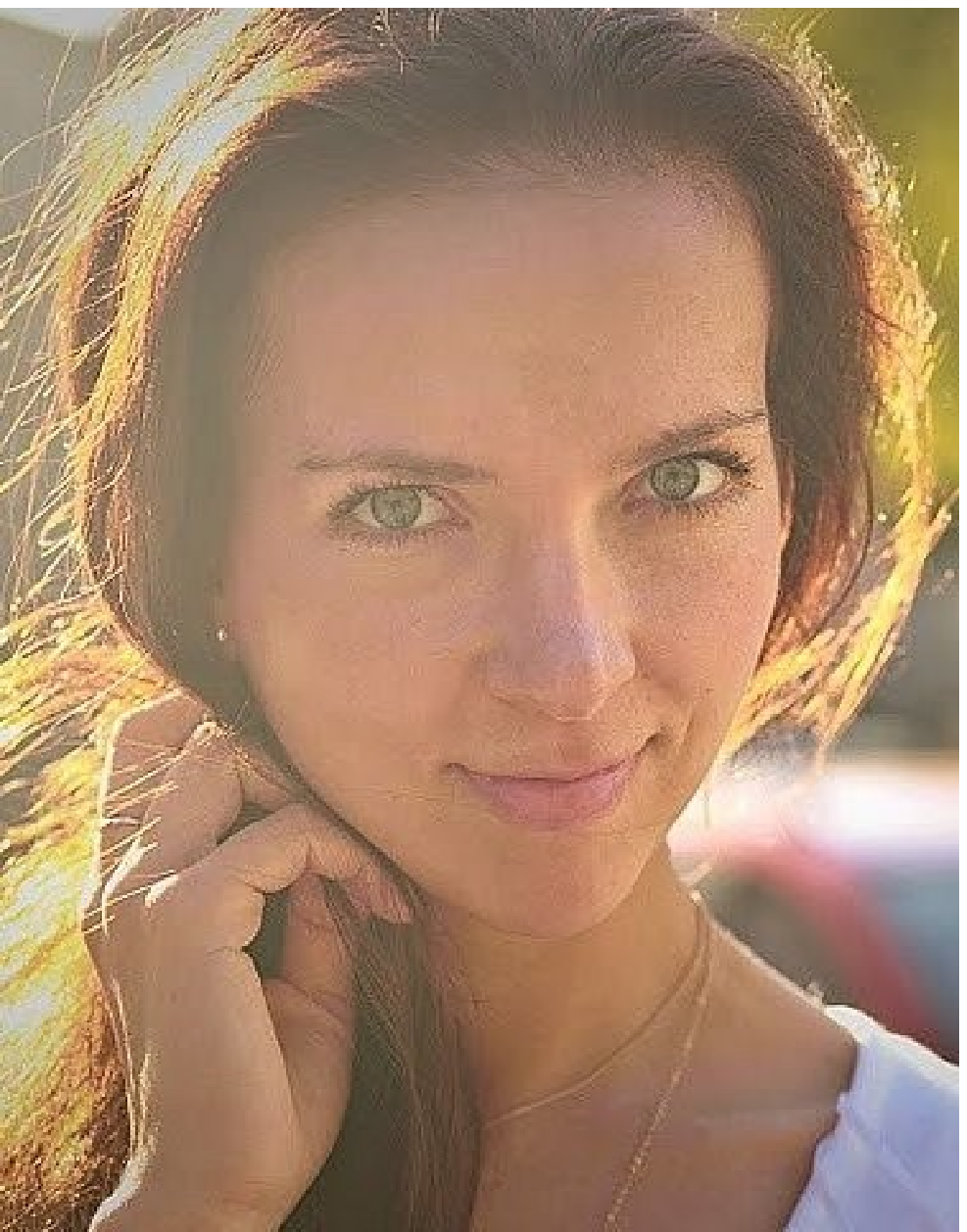}}]{Olga Ganinina}
		is a Research Scientist in the Department of Electronics and Communications Engineering at Tampere University of Technology, Finland. She received her B.Sc. and M.Sc. degrees in Applied Mathematics from Department of Applied Mathematics, Faculty of Mechanics and Physics, St. Petersburg State Polytechnical University, Russia as well as the Ph.D. degree from Tampere University of Technology. Her research interests include applied mathematics and statistics, queueing theory and its applications; wireless networking and energy efficient systems, machine-to-machine and device-to-device communication.
\end{IEEEbiography}
\vspace{-1.5cm}
\begin{IEEEbiography}[{\includegraphics[width=1in,height=1.25in,clip,keepaspectratio]{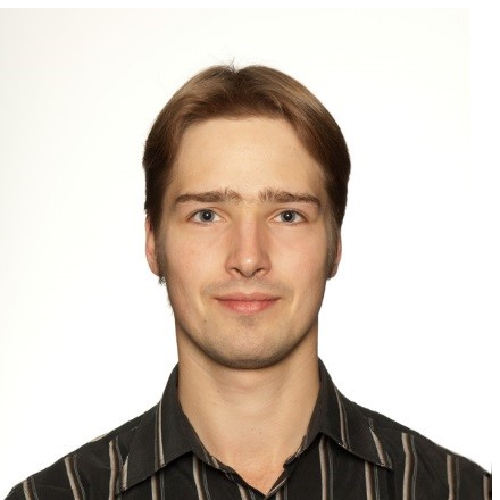}}]{Sergey Andreev}
		is a Senior Research Scientist in the Department of Electronics and Communications Engineering at Tampere University of Technology, Finland. He received the Specialist degree (2006) and the Cand.Sc. degree (2009) both from St. Petersburg State University of Aerospace Instrumentation, St. Petersburg, Russia, as well as the Ph.D. degree (2012) from Tampere University of Technology. Sergey (co-)authored more than 90 published research works on wireless communications, energy efficiency, heterogeneous networking, cooperative communications, and machine-to-machine applications.
\end{IEEEbiography}

\begin{IEEEbiography}[{\includegraphics[width=1in,height=1.25in,clip,keepaspectratio]{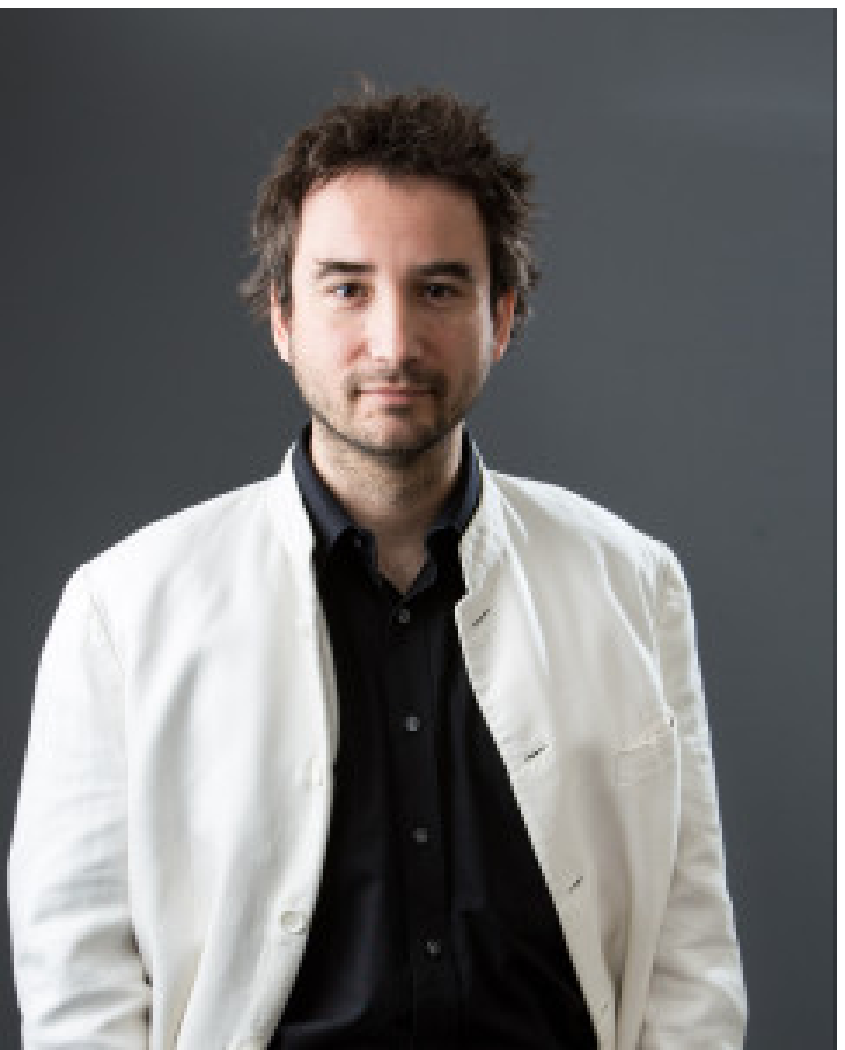}}]{Mischa Dohler}
		 is a Full Professor of Wireless Communications with King’s College London, London, U.K., the Head of the Centre for Telecommunications Research, the Co-Founder and a Member of the Board of Directors of the smart city pioneer Worldsensing. He is a frequent keynote, panel, and tutorial speaker. He has pioneered several research fields, contributed to numerous wireless broadband, IoT/M2M, and cyber security standards, holds a dozen patents, organized, and chaired numerous conferences, has more than 200 publications, and authored several books. He acts as policy, technology, and entrepreneurship adviser. He has talked at TEDx and has had TV \& radio coverage.
He is a Distinguished Lecturer of the IEEE. He is the Editor-in-Chief of the \emph{Transactions on Emerging Telecommunications Technologies (Wiley)} and the \emph{EAI Transactions on the Internet of Things}.
\end{IEEEbiography}
\vfill
\end{document}